\documentclass[aps,prd,notitlepage,nofootinbib,showpacs,10pt,tightenlines]{revtex4-1}

\usepackage{amssymb,amsmath,bm}
\usepackage{hyperref} 

\usepackage{graphics, graphicx}
\usepackage{color}

\newcommand{\be}{\begin{equation}}
\newcommand{\ee}{\end{equation}}
\newcommand{\bea}{\begin{eqnarray}}
\newcommand{\eea}{\end{eqnarray}}
\newcommand{\bem}{\begin{multline}}
\newcommand{\eem}{\end{multline}}
\newcommand{\beg}{\begin{gather}}
\newcommand{\eeg}{\end{gather}}

\newcommand{\as}{\alpha_s}

\def\eq#1{{Eq.~(\ref{#1})}}
\def\fig#1{{Fig.~\ref{#1}}}
\newcommand{\ben}{\begin{eqnarray*}}
\newcommand{\een}{\end{eqnarray*}}

\def\peq#1{{(\ref{#1})}}

\begin{document}

\title{A New Mechanism for Generating a Single Transverse Spin
  Asymmetry} 

\author{Yuri~V.~Kovchegov,\footnote{kovchegov.1@asc.ohio-state.edu}
  Matthew D. Sievert\footnote{sievert.7@osu.edu}}

\affiliation{Department of Physics, The Ohio State University, Columbus, OH 43210, USA}

\begin{abstract}
  We propose a new mechanism for generating a single transverse spin
  asymmetry (STSA) in polarized proton--proton and proton--nucleus
  collisions in the high-energy scattering approximation. In this
  framework the STSA originates from the $q \to q \, G$ splitting in
  the projectile (proton) light-cone wave function followed by a
  perturbative ($C$-odd) odderon interaction, together with a $C$-even
  interaction, between the projectile and the target. We show that
  some aspects of the obtained expression for the STSA of the produced
  quarks are in qualitative agreement with experiment: STSA decreases
  with decreasing projectile $x_F$ and is a non-monotonic function of
  the transverse momentum $k_T$. In our framework the STSA peaks at
  $k_T$ near the saturation scale $Q_s$. Our mechanism predicts that
  the quark STSA in proton--nucleus collisions should be much smaller
  than in proton--proton collisions.  We also observe that in our
  formalism the STSA for prompt photons is zero.
\end{abstract}


\maketitle


\section{Introduction}
\label{sec-Intro}

\begin{figure}[h]
  \includegraphics[width=5cm]{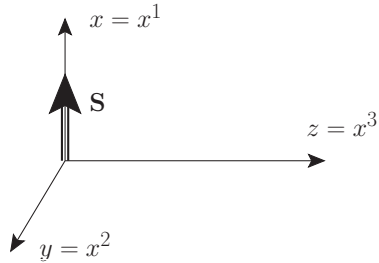}
 \caption{A sketch of the geometry of polarized scattering: the incoming 
   transversely-polarized projectile is moving along the $+z$-axis,
   while its spin is pointing along the $+x$-axis. The positive
   $y$-axis points to the right of the beam.}
\label{orientation} 
\end{figure}

In recent years a significant theoretical effort has been directed
toward understanding the origin and the interesting properties of
transverse spin asymmetries observed in high-energy scattering
experiments. The \emph{single transverse spin asymmetry} (STSA) in
polarized scattering $P^{\uparrow} (p) + A \rightarrow h (k) + X$ is
an observable describing the correlation between the transverse spin
vector $\bm S$ of the projectile $P$ and the transverse momentum $\bm
k$ of the produced hadron $h$.  This correlation can be expressed
either in terms of the asymmetry in the scattering of the
spin-up and spin-down 
transverse spin states, or in terms of the ``left-right'' ($\bm k
\leftrightarrow - \bm k$) asymmetric momentum distribution in spin-up
scattering:
\begin{equation}
 \label{eq-Defn STSA}
   A_N ({\bm k}) \equiv \; \frac{\frac{d \sigma^\uparrow}{d^2 k \, dy} - 
   \frac{d \sigma^\downarrow}{d^2 k \, dy}} {\frac{d \sigma^\uparrow}{d^2 k \, dy} + 
   \frac{d \sigma^\downarrow}{d^2 k \, dy}} \; = \;
   \frac{\frac{d \sigma^\uparrow}{d^2 k \, dy} (\bm k) - 
   \frac{d \sigma^\uparrow}{d^2 k \, dy} (- \bm k)}
   {\frac{d \sigma^\uparrow}{d^2 k \, dy} (\bm k) + 
   \frac{d \sigma^\uparrow}{d^2 k \, dy} (- \bm k)}
   \; \equiv \frac{d(\Delta \sigma)}{2 \, d\sigma_{unp}}.
\end{equation}
Interpreted as a left-right asymmetry in the produced hadron
distribution, the sign of $A_N$ is fixed to reflect a preferential
scattering of particles to the beam-left when $A_N$ is positive and a
preferential scattering to the beam-right when $A_N$ is negative.
More concretely, if the polarized projectile moves along the $+z$ axis
and the transverse spin is oriented along the $+x$ axis, then positive
$A_N$ corresponds to more outgoing particles produced along the $-y$
axis than along the $+y$ axis. The geometry of polarized scattering is
illustrated in \fig{orientation}. The STSA can also be thought of as a
spin-momentum coupling term proportional to $({\vec S} \times {\vec p}
\, ) \cdot {\vec k}$ in the particle production cross section, where
$\vec p$ is the momentum of the incoming projectile.  For a review of
the current status of STSA physics see \cite{D'Alesio:2007jt}.

\begin{figure}
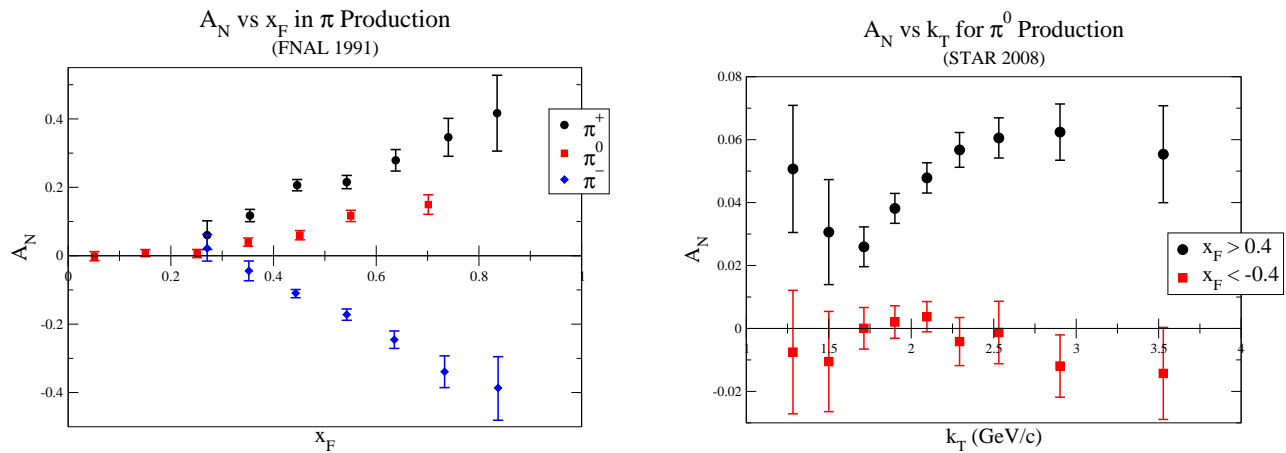

 \begin{tabular}{c c}
 \includegraphics[width=8cm]{FNAL_91_xF_Improved2.eps} \hspace*{3mm} & \hspace*{3mm}
 \includegraphics[width=8cm]{STAR_08_pT.eps}
 \end{tabular}
  \caption{Experimental data on the pion single transverse spin asymmetry 
    $A_N$ as a function of Feynman-$x$ reported by E581 and E704
    collaborations (graphically reconstructed from
    \cite{Adams:1991cs}, shown in the left panel) for $0.7 \le k_T \le
    2.0$~GeV/c, and as a function of the pion transverse momentum
    $k_T$ collected by the STAR collaboration \cite{Abelev:2008qb}
    (right panel).}
 \label{fig-Experimental Data}  
\end{figure}

In the 1990's, the E581 and E704 collaborations at Fermilab reported
data on hadron production in transversely polarized proton-proton
scattering that showed large unanticipated transverse spin asymmetries
of up to 30-40\%
\cite{Adams:1991rw,Adams:1991cs,Adams:1991ru,Adams:1995gg,Bravar:1996ki,Adams:1994yu}.
Qualitatively, the asymmetries were consistent with zero for mid
rapidities, but increased rapidly in the forward scattering direction,
as illustrated by the data shown in the left panel of
\fig{fig-Experimental Data}.  More recently, the PHENIX, STAR, and
BRAHMS collaborations at the Relativistic Heavy Ion Collider (RHIC)
have studied transverse spin asymmetries over a wide kinematic range
at $\sqrt{s} = 200 \, GeV$
\cite{Abelev:2007ii,Nogach:2006gm,Adler:2005in,Lee:2007zzh}.  The data
they have presented \cite{Abelev:2007ii,Nogach:2006gm,Adler:2005in}
confirmed and extended the Fermilab results, and also indicated a
non-monotonic dependence of STSA on the transverse momentum $k_T$ of
the produced hadron \cite{Abelev:2008qb,Wei:2011nt}, shown in the
right panel of \fig{fig-Experimental Data}.

At the time these spin asymmetries were first observed, there was no
theoretical framework to understand them; on the contrary, prevailing
wisdom expected spin-dependent effects to become negligible at high
energies \cite{Kane:1978nd}.  Much theoretical progress in classifying
and modeling sources of STSA has been made since then, predominantly
within the framework of collinear factorization.

\begin{figure}[b]
\includegraphics[width=10cm]{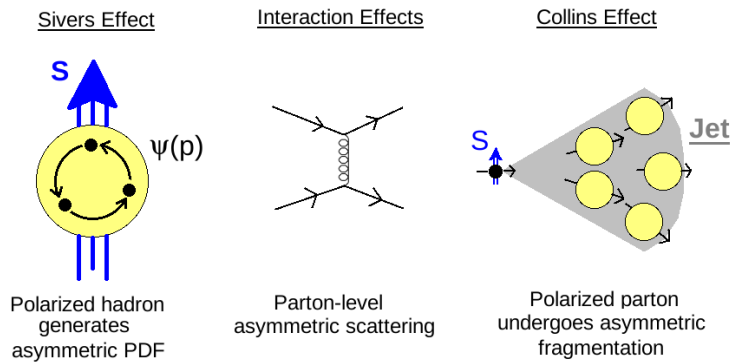}
 \caption{Sketch of the three potential sources of 
   asymmetry: in the parton distribution function via the Sivers
   effect, via partonic interaction, or in the fragmentation function
   via the Collins effect.}
\label{fig-Mechanisms}  
\end{figure} 

There are three stages in the scattering process where an asymmetry
could be generated, illustrated in \fig{fig-Mechanisms}.  Sivers
\cite{Sivers:1989cc,Sivers:1990fh} proposed the existence of a
correlation between the spin of the polarized hadron and the
transverse momentum distribution of its partons.  This Sivers effect
would act within the parton distribution function (PDF) (or,
equivalently, the initial state wave function) of the polarized
projectile to generate an asymmetry. Being a part of the PDF, this is
an intrinsically non-perturbative process. The asymmetry could also be
generated in the partonic scattering processes themselves; since the
leading-order scattering process does not generate an asymmetry, any
contribution would come from ``higher-twist'' terms in the
interaction, as suggested in
\cite{Efremov:1981sh,Efremov:1984ip,Qiu:1991pp,Ji:1992eu,Qiu:1998ia,Brodsky:2002cx,Collins:2002kn,Koike:2011mb,Kanazawa:2000hz,Kanazawa:2000kp}.
Such an asymmetry generated by the interaction may be perturbative,
depending on the kinematics.  Finally, Collins \cite{Collins:1992kk}
proposed a similar correlation between spin and transverse momentum
that could occur during hadronization, as a produced parton undergoes
fragmentation into final-state hadrons.  The Collins effect would
couple the spin of a produced parton to the momentum dependence of its
fragmentation function, resulting in a left-right asymmetry within the
final-state jet. Again, the Collins effect is a low-energy,
intrinsically non-perturbative process.

\begin{figure}[h]
 \includegraphics[width=8cm]{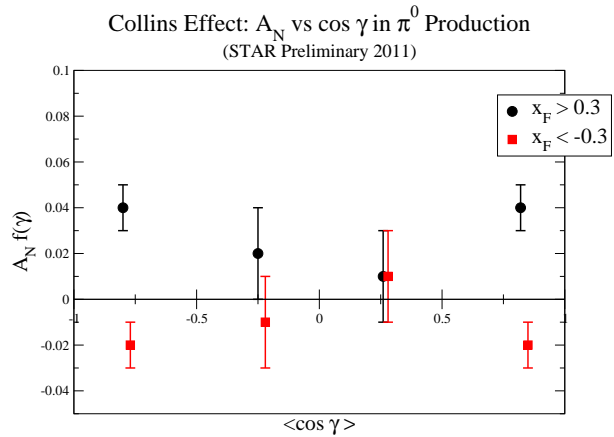}
 \caption{Recent preliminary data from the STAR collaboration (graphically reconstructed 
   from \cite{Poljak:2010tm}) for the $\pi^0$ asymmetry within
   final-state jets as a function of the angle $\gamma$ between the
   outgoing pion and the jet thrust axis.  The Collins contribution is
   proportional to the slope of the data, and is consistent with
   zero.}
\label{fig-Collins}  
\end{figure}

Recent studies from STAR \cite{Poljak:2010tm} were able to resolve the
angular distribution within the jets and thus probe the Collins
function directly.  Their data suggests that the Collins effect's
contribution to the asymmetry is small or consistent with zero, as
illustrated in Fig.~\ref{fig-Collins}.

We are interested in understanding which perturbative mechanisms
within the interaction can give rise to transverse spin asymmetries,
and we would like to study the interaction with proton--nucleus
($p^\uparrow + A$) collisions in mind.  When the target is a heavy
nucleus at high energies, or when the energy of the proton--proton
collision is high enough, the projectile scatters off of the target's
small-$x$ parton distribution where parton densities are high.  The
effective interactions in this small-$x$ regime are well described by
the parton saturation/Color Glass Condensate (CGC) formalism
\cite{Gribov:1984tu},\cite{Blaizot:1987nc},\cite{Mueller:1986wy},\cite{Mueller:1994rr,Mueller:1994jq,Mueller:1995gb},\cite{McLerran:1993ka,McLerran:1993ni,McLerran:1994vd},\cite{Kovchegov:1996ty,Kovchegov:1997pc},\cite{Jalilian-Marian:1997xn,Jalilian-Marian:1997jx,Jalilian-Marian:1997gr,Jalilian-Marian:1997dw,Jalilian-Marian:1998cb,Kovner:2000pt,Weigert:2000gi,Iancu:2000hn,Ferreiro:2001qy},\cite{Balitsky:1996ub,Balitsky:1997mk,Balitsky:1998ya,Kovchegov:1999yj,Kovchegov:1999ua},\cite{Iancu:2003xm,Weigert:2005us,Jalilian-Marian:2005jf}
which resums the multiple rescatterings in these dense color fields
and incorporates the small-$x$ evolution of those fields.  This
saturation framework is expressed naturally in coordinate space and
introduces a characteristic saturation scale $Q_s$ describing the
color-charge density fluctuations within the target.

With the CGC interactions that mediate $p^\uparrow + A$ scattering in
mind, it is natural to search for the asymmetry using the language of
the small-$x$ interactions, light-cone perturbation theory (LCPT)
\cite{Lepage:1980fj,Brodsky:1997de}, instead of the usual collinear
factorization.  Using LCPT, we calculate spin-dependent processes that
contribute to the projectile's light-cone wave function, then
convolute this wave function (squared) with the known scattering
amplitudes from CGC that are responsible for the interaction of
partons with the target. Surprisingly enough this procedure yields a
non-zero STSA for quark and gluon production in $p^\uparrow + A$
scattering. Our approach yields a different but complementary picture
to the usual collinear factorization mechanisms for generating STSA
and may provide new insight into the underlying physics for generating
transverse spin asymmetry.

In the past the interplay of spin and small-$x$ evolution was
investigated in \cite{Bartels:1995iu} using the standard Feynman
diagram approach. In more recent years the role of spin has also been
investigated within the McLerran-Venugopalan (MV) model
\cite{McLerran:1993ka,McLerran:1993ni,McLerran:1994vd} in
\cite{Metz:2011wb} and within Mueller's dipole model
\cite{Mueller:1994rr,Mueller:1994jq,Mueller:1995gb} in
\cite{Dominguez:2011br}. The transverse spin asymmetry in the CGC
formalism was studied in
\cite{Boer:2006rj,Boer:2002ij,Boer:2008ze,Kang:2011ni}, with the STSA
being a completely non-perturbative effect on top of the perturbative
CGC dynamics. Other non-perturbative approaches include using QCD
instantons to generate STSA as discussed in \cite{Qian:2011ya}.

In this paper, we present an analysis of STSA generated by the
interaction of the spin-dependent LCPT wave function of the projectile
with the target gluon field in the saturation/CGC framework.  The
paper is organized as follows.  In Sec.~\ref{sec-General Result} we
derive the general expressions for the quark, gluon, and photon single
transverse spin asymmetries in our formalism. These general results
are given in Eqs.~\peq{eq-observables}, \peq{direct3} and \peq{Gprod}.
We show that to generate the asymmetry the projectile needs to
interact with the target via a $C$-odd scattering amplitude, commonly
known as the QCD odderon
\cite{Lukaszuk:1973nt,Nicolescu:1990ii,Ewerz:2003xi,Bartels:1999yt,Kovchegov:2003dm,Hatta:2005as,Kovner:2005qj,Jeon:2005cf}.
In Sec.~\ref{sec-Estimates} we simplify the interaction of the
light-cone wave function with the target using the Glauber-Mueller
approximation \cite{Mueller:1989st}, and we present a closed-form
estimate of the asymmetry for produced quarks. We find that a $C$-odd
exchange alone is insufficient to generate the asymmetry, and needs to
be accompanied by a $C$-even exchange for the asymmetry to be
non-zero. We also see that the generated STSA ($A_N$) is a
non-monotonic function of transverse momentum $k_T$ and an increasing
function of $x_F$, in qualitative agreement with the data shown in
\fig{fig-Experimental Data}. Our rough estimates of the asymmetry are
plotted in Figs.~\ref{AN}, \ref{AN2} and \ref{ANrad}, which, when
compared to the data in Fig.~\ref{fig-Experimental Data}, suggest that
it may even be possible to achieve quantitative agreement with the
data using our approach in a detailed phenomenological study.  We
conclude by summarizing our main results in Sec.~\ref{sec-Concl}.



\section{General Result: Coupling Spin to Interaction C-Parity}

\label{sec-General Result} 

\begin{figure}[h]
\includegraphics[width=8cm]{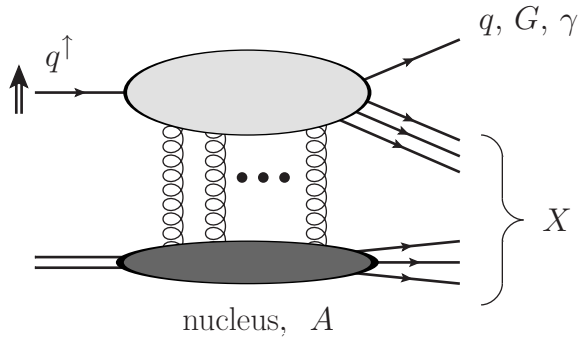}
 \caption{Transversely polarized quark scattering in the field of the 
   nucleus producing either a quark ($q$), gluon ($G$), or a prompt
   photon ($\gamma$) along with extra hadrons denoted by $X$:
   $q^{\uparrow} +  A \rightarrow (q, \, G, \, \gamma) + X$.} 
\label{fig-QuarkScattering}  
\end{figure}

In this paper we consider two processes contributing to the STSA of
hadron production in the scattering of transversely polarized protons
on an unpolarized target (proton or nucleus), $p^{\uparrow} + A
\rightarrow h + X$: quark production and gluon production. We will
also consider STSA in prompt photon production, $p^{\uparrow} + A
\rightarrow \gamma + X$. The proton's transverse spin is transmitted
to a quark in its wave function (giving rise to the transversity
distribution); for simplicity we consider the scattering of this
transversely polarized quark on the target separately from other
(spectator) quarks in the proton. This process is illustrated in
\fig{fig-QuarkScattering}, where the high energy interaction between
the projectile quark and the target nucleus is schematically denoted
by gluon exchanges.

\begin{figure}[h]
\includegraphics[width=15cm]{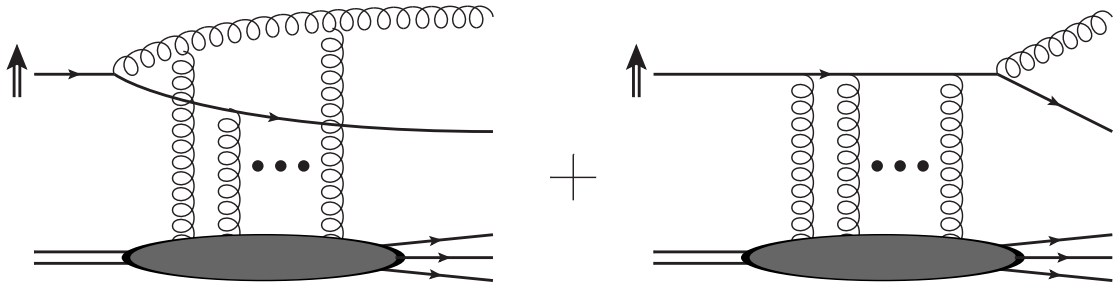}
 \caption{Two contributions to the amplitude for the high energy 
   quark--target scattering in LCPT.}
\label{qtoqGampl}  
\end{figure}

To introduce the methodology, we will concentrate on the quark
production process, $q^{\uparrow} + A \rightarrow q + X$.  We are
interested in isolating the spin-dependent contribution of the
quark--target scattering. Certainly keeping only the eikonal
interaction would be insufficient to reach this goal, since the
eikonal scattering is independent of the quark polarization. A
non-eikonal correction has to be included: in the
multiple-rescattering Glauber-Mueller \cite{Mueller:1989st}
approximation, the non-eikonal rescattering corrections are suppressed
by powers of energy and are very small. A much larger spin-dependent
contribution comes from the non-eikonal splitting of the projectile
quark into a quark and a gluon, $q \to q \, G$, which is suppressed
only by a power of the strong coupling $\as$. In the LCPT language the
$q \to q \, G$ splitting may take place either before or after the
interaction with the target, as shown in \fig{qtoqGampl}. (Splitting
during the interaction with the target is suppressed by powers of
energy \cite{Kovchegov:1998bi}.) The lowest-order diagrams shown in
\fig{qtoqGampl} that contribute to STSA in $q^{\uparrow} + A
\rightarrow q + X$ contain the emission of a single gluon from the
polarized quark, where both the gluon and quark can scatter in the
field of the target. (Multi-gluon non-eikonal emissions are also
possible, but they are higher-order in $\as$ and, hence, outside of
the leading-order precision of this work.)  The spin dependence of the
process illustrated in \fig{qtoqGampl} originates within the
light-cone wave function of the quark-gluon system, which couples to
the interaction in a way that generates the asymmetry. In this
Section, we will first outline the calculation of the $q \to q \, G$
light-cone wave function using LCPT.  Then we will combine the
resulting splitting wave function squared with the quark and gluon
interactions in the field of the target and identify the contribution
to the asymmetry. In the end we obtain general expressions for quark,
gluon, and photon STSA's in our formalism. The incoming light quark
has a particular flavor $f$; multiple quark flavors can be
incorporated into our formalism by convoluting the obtained cross
sections with quark distributions corresponding to different flavors
(inserting the appropriate quark masses into our results below).

Throughout this paper, we will work in light-cone coordinates $p^\mu =
(p^+ , p^- , \bm p )$ with $p^\pm \equiv p^{0} \pm p^{3}$ and the
transverse-plane vector ${\bm p} \equiv ( p^{1} , p^{2} )$. Note that
$p \cdot q = (1/2) \, ( p^+ q^- + p^- q^+ ) - \bm p \cdot \bm q$ and
$p_\mu \, p^\mu = p^+ p^- - p_T^2$ with $p_T = |{\bm p}|$. We assume
that the incoming projectile quark is moving along the light-cone
$x^+$-axis, while the target is moving along the $x^-$-axis. We will
work in the light-cone gauge of the projectile, $A^+ = 0$.



\subsection{Light-Cone Wave Function and Transverse Polarization}

\label{subsec-Wavefn}

\begin{figure}[h]
\includegraphics[width=8cm]{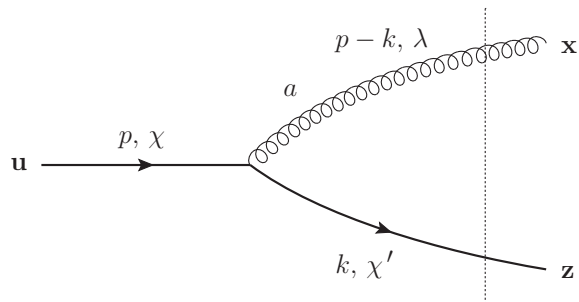}
\caption{The light-cone wave function for the $q \to q \,
  G$ splitting. Vertical dotted line denotes the intermediate state.}
\label{fig-SplittingVertex}  
\end{figure}

Consider the splitting shown in \fig{fig-SplittingVertex} of a
transversely polarized quark with momentum $p$ and polarization $\chi
= \pm 1$ decaying into a gluon (with momentum $p-k$, polarization
$\lambda$, and color $a$) and a recoiling quark (with momentum $k$ and
polarization $\chi'$).  The projectile quark is traveling along the
light-cone $x^+$-direction and the recoiling quark carries a fraction
\begin{equation}
 \label{eq-momentum fraction}
 \alpha \equiv \frac{k^+}{p^+}
\end{equation}
of the incoming quark's longitudinal momentum.  We do not restrict
ourselves to the case of an eikonal quark emitting a soft gluon ($1 -
\alpha \ll 1$), but work in the general case when both the quark and
the gluon can carry comparable longitudinal momenta.

The calculation of the light-cone wave function corresponding to the
diagram in \fig{fig-SplittingVertex} is different from other similar
calculations in the literature (see e.g.
\cite{Itakura:2003jp,Albacete:2006vv}) only in that now the incoming
quark is polarized {\sl transversely}. To account for the transverse
spin of the projectile traveling in the $x^{3} = z$ direction, we
define the axis of spin quantization to be the $x^{1} = x$ axis, and
we need to construct the Dirac spinors corresponding to the spin
projection eigenvalues $\chi = \pm 1$ on the $x$-axis.  The proper
Lorentz-covariant operator describing the spin orientation is the
Pauli-Lubanski vector
\begin{equation}
 \label{eq-Pauli Lubanski}
 W_\mu \equiv - \frac{1}{2} \epsilon_{\mu \nu \rho \sigma} S^{\nu \rho} p^\sigma \;,
\end{equation}
where
\begin{equation}
 \nonumber
 S^{\nu \rho} \equiv \frac{i}{4} \big[ \gamma^\nu \, , \, \gamma^\rho \big]
\end{equation}
are the generators of the Lorentz group and we are using the
convention in which $\epsilon_{0123} = +1$.

For our purposes, we need to find the eigenspinors of $W_{1}$ for a
particle moving along the $z$-axis.  Diagonalizing $W_{1}$ in terms of
the pure helicity eigenspinors gives (cf. e.g. \cite{Cortes:1991ja})
\begin{equation}
 \label{eq-spinors}
 U_\chi \equiv \frac{1}{\sqrt{2}} \big[ U_{(+z)} + \chi \, U_{(-z)} \big] \; ,
\end{equation}
where $U_{\pm z}$ are the spinors in the helicity basis and $\chi =
\pm 1$. These particular spinors $U_\chi$ are simultaneous eigenstates
of $W_{1}$ and of the Dirac operator:
\begin{eqnarray}
 \nonumber
 W_{1} \, U_\chi = \chi \, \frac{m}{2} \, U_\chi  \\ \nonumber
 (\gamma \cdot p - m) \, U_\chi = 0 \;.
\end{eqnarray}

Since the incoming quark is polarized transversely, it is convenient
to use the same spinor basis \peq{eq-spinors} for the outgoing quark
in \fig{fig-SplittingVertex} as well. Using the standard rules of LCPT
\cite{Lepage:1980fj,Brodsky:1997de} and working in the transverse
polarization basis \peq{eq-spinors} for the spinors, we evaluate the
light-cone wave function shown in \fig{fig-SplittingVertex} as
\begin{equation}
 \label{wf1}
 \psi_{\lambda \chi \chi'}^a ({\bm k}, {\bm p}, \alpha) =
 \frac{g \, T^a}{p^- - k^- - (p-k)^-} \;
 \bigg[ \frac{\bar U_{\chi'} (k)}{\sqrt{k^+}} \, \gamma \cdot \epsilon_{\lambda}^* \,
 \frac{U_\chi (p)}{\sqrt{p^+}} \bigg] \; ,
\end{equation}
where 
\begin{equation}
  \label{eq:pol}
  \epsilon_\lambda^\mu = \left( 0, \frac{2 \, {\bm \epsilon}_\lambda \cdot
 ({\bm p} -{\bm k})}{p^+ - k^+}, {\bm \epsilon}_\lambda \right)
\end{equation}
is the gluon polarization vector with the transverse components ${\bm
  \epsilon}_\lambda = (-1/\sqrt{2}) \, (\lambda, i)$.

In arriving at \eq{wf1} we have used the fact that the incoming state
in \fig{fig-SplittingVertex} contains only the quark with momentum $p$
while the intermediate state contains the quark and the gluon, as
denoted by the vertical dotted line in \fig{fig-SplittingVertex}. For
diagrams where the polarized quark scatters in the nucleus before the
gluon emission, as shown in the right panel of \fig{qtoqGampl}, the
roles are reversed: the quark line $p$ is the intermediate state, and
the quark--gluon system is the final state.  Since $\sum_{init} p_i^-
= \sum_{final} p_i^-$, the energy denominator reverses sign for
final-state splittings.

Using the on-shell conditions explicitly gives the terms entering the
energy denominator:
\begin{equation}
 \label{eq-minus components}
 p^- = \frac{p_T^2 + m^2}{p^+}  \;\;,\;\;
 k^- = \frac{k_T^2 + m^2}{k^+} \;\;,\;\;
 (p-k)^- = \frac{(\bm p - \bm k)^2}{p^+ - k^+} \;\; .
\end{equation}
We use the spinors in the Brodsky--Lepage convention
\cite{Lepage:1980fj,Brodsky:1997de}, which, for a particle moving
along the $z$-axis, become helicity eigenstates.  The spinor matrix
elements for the Brodsky--Lepage spinors are well known
\cite{Lepage:1980fj,Brodsky:1997de}.  Making the change of basis
\eqref{eq-spinors} gives the relevant spinor products as
\footnote{Note again that $U_\chi (p)$ becomes a spinor for a
  transversely polarized particle only for $\bm p =0$:
  Eqs.~\peq{eq-transverse matrix elts} give us the matrix elements for
  spinors related to the Brodsky--Lepage spinors via \eq{eq-spinors},
  which do not necessarily correspond to transverse polarizations in
  the general case.}
\begin{subequations}
 \label{eq-transverse matrix elts}
 \begin{eqnarray}
  \frac{\bar U_{\chi'} (k)}{\sqrt{k^+}} \, \gamma^+ \, \frac{U_\chi (p)}{\sqrt{p^+}} &=&
  2 \, \delta_{\chi, \chi'} \\
  \frac{\bar U_{\chi'} (k)}{\sqrt{k^+}} \, \gamma_\bot^i \, \frac{U_\chi (p)}{\sqrt{p^+}} &=&
  \frac{\delta_{\chi,\chi'}}{\alpha \, p^+} \bigg[ 
   (k_\perp^i + \alpha \, p_\perp^i) + (1-\alpha) \, i \, m \, \chi \, \delta^{i2}
  \bigg] \\ \nonumber
  &-& \frac{\delta_{\chi,-\chi'}}{\alpha \, p^+} \bigg[
   i \, \epsilon^{ij} \, (k^j_\perp - \alpha \, p^j_\perp) + (1-\alpha) \, m \, \chi \, \delta^{i1}
  \bigg] \;,
 \end{eqnarray}
\end{subequations}
and the $\gamma^-$ matrix element does not contribute to $\gamma \cdot
\epsilon^*_\lambda$ since $\epsilon_\lambda^+ = 0$ in the light-cone
gauge. Here $\epsilon^{12} = - \epsilon^{21} =1$, $\epsilon^{11} =
\epsilon^{22} =0$. With the matrix elements \peq{eq-transverse matrix
  elts} it is straightforward to evaluate the light-cone wave function
\eqref{wf1} in momentum space, obtaining
\begin{eqnarray}
 \psi_{\lambda \chi \chi'}^a ( {\bm k} , {\bm p}, \alpha) &=& \frac{g \, T^a}{({\bm k} -   
 \alpha \, {\bm p})^2 + {\tilde m}^2} \notag \\ & \times &
 \bigg[ {\bm \epsilon}_{\lambda}^* \cdot ({\bm k} - \alpha \, {\bm p}) \, \bigg(
 (1+\alpha) \, \delta_{\chi \chi'} + \lambda \, (1-\alpha) \, \delta_{\chi, -\chi'} \bigg) 
 - \frac{\tilde m}{\sqrt 2} \, (1-\alpha) \, \chi \,
 \big (\delta_{\chi \chi'} - \lambda \, \delta_{\chi, -\chi'} \big) \bigg] \;, 
 \label{eq-momentumwavefn}
\end{eqnarray}
where
\begin{equation}
 \nonumber
 \tilde m \equiv (1-\alpha) m
\end{equation}
is a natural effective mass parameter in the wave function and $T^a$
are the SU($N_c$) generators in the fundamental representation.

Now we can Fourier transform the wave function to coordinate space
\begin{equation}
 \label{eq-Fourier transform}
 \psi_{\lambda \chi \chi'}^a ({\bm x}, {\bm z}, \alpha; {\bm u} ) \equiv 
 \int \frac{d^2 k}{(2\pi)^2} \frac{d^2p }{(2\pi)^2} e^{i \, {\bm k} \cdot ({\bm z} -
 {\bm x})} \, e^{i \, {\bm p} \cdot ({\bm x - \bm u})} \,
 \psi_{\lambda \chi \chi'}^a ({\bm k}, {\bm p}, \alpha) \; 
\end{equation}
with the transverse coordinates defined in \fig{fig-SplittingVertex}.
Since the momentum-space wave function depends only on $\bm k - \alpha
\, \bm p$, one of the two integrals can be performed to yield a delta
function $\delta^2 [(\bm x - \bm u) + \alpha \, (\bm z - \bm x)]$.
Performing the remaining momentum integral in \eq{eq-Fourier
  transform} yields
%
%
\begin{eqnarray}
 \psi_{\lambda \chi \chi'}^a ({\bm x}, {\bm z}, \alpha ; {\bm u}) &=& 
\frac{g \, T^a}{2\pi} \, \delta^2 [({\bm x} - {\bm u}) + \alpha ({\bm z - \bm x}) ]  
 \, {\tilde m}  \, \bigg\{ 
  i \, {\bm \epsilon}_\lambda^* \cdot \frac{{\bm z - \bm x}}{{|\bm
 z - \bm x|}} \, 
 K_1 ( \tilde m \, |\bm z - \bm x|) \bigg[ (1+\alpha) \, \delta_{\chi,\chi'} +
 \lambda \, (1-\alpha) \, \delta_{\chi, -\chi'} \bigg] \nonumber \\
 &-& \frac{\chi \, (1-\alpha)}{\sqrt 2} \, K_0 (\tilde m \, |\bm z - \bm x|) \, \bigg[ \delta_{\chi, \chi'} - \lambda \,
 \delta_{\chi, - \chi'} \bigg] \bigg\}. \label{eq-coord wavefn}
\end{eqnarray} 
It is useful to separate out the color factor $T^a$ and the delta
function from the rest of the wave function (denoted by $\Psi_{\lambda
  \chi \chi'}$), such that
\begin{equation}
 \label{eq-Psi vs Phi}
 \psi_{\lambda \chi \chi'}^a (\bm x , \bm z , \alpha ; \bm u) \equiv
 T^a \, \delta^2 [(\bm u - \bm x) - \alpha \, (\bm z - \bm x)] \,
 \Psi_{\lambda \chi \chi'} (\bm z - \bm x, \alpha) \;.
\end{equation}

\begin{figure}[h]
\includegraphics[width=12cm]{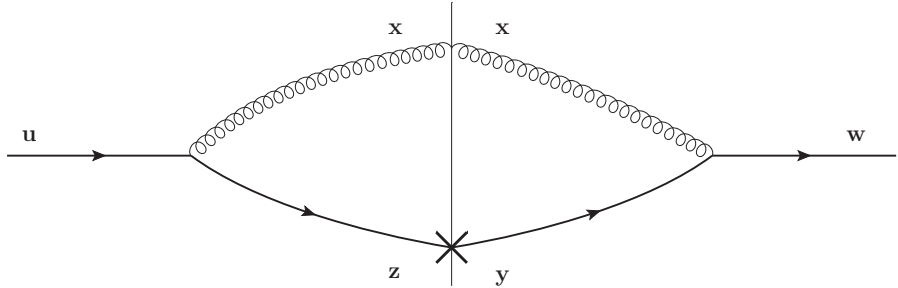}
 \caption{Light-cone wave function from \fig{fig-SplittingVertex} squared. 
   The vertical straight line separates the wave function from its
   conjugate, while the cross denotes the quark that we tag on. The
   untagged gluon's coordinate $\bm x$ is unchanged, but the quark
   coordinates differ ($\bm z$ and $\bm u$ vs. $\bm y$ and $\bm w$, as
   explained in the text.)}
\label{fig-wavefn squared}  
\end{figure}

Finally, we need to square the wave function and sum over the final
particles' polarizations. Here we are interested in producing a quark
with a fixed transverse momentum, while integrating over all
transverse momenta of the produced gluon in \fig{qtoqGampl}. According
to the standard prescription
\cite{Kovchegov:1998bi,Jalilian-Marian:2005jf}, for coordinate-space
scattering amplitudes this means that the gluon's transverse
coordinate $\bm x$ will be the same both in the amplitude and in the
complex conjugate amplitude (since its transverse momentum is
integrated over in the cross section), while the quarks have different
transverse coordinates between the amplitude and the conjugate
amplitude (since they are the observed particles).  (See
\fig{fig-amplitude squared} below for the illustration of the
amplitude squared.) The ``square'' of the light cone wave function
\peq{eq-coord wavefn} with the above rule for the quark and gluon
transverse coordinates is illustrated in \fig{fig-wavefn squared}.
     
The wave function \peq{eq-coord wavefn} squared $\Phi_\chi$ as shown
in \fig{fig-wavefn squared} contains one contribution which is
polarization-independent and another which is proportional to the
quark polarization eigenvalue $\chi$
\begin{equation}
 \label{eq-wavefn squared}
 \Phi_\chi (\bm z - \bm x , \bm y - \bm x, \alpha) \equiv \sum_{\lambda \, , \, \chi' = \pm 1} \, 
 \Psi_{\lambda \chi \chi'} (\bm z - \bm x, \alpha) \, \Psi^{*}_{\lambda \chi \chi'} 
 (\bm y - \bm x, \alpha) \, \equiv \, \Phi_{unp} (\bm z - \bm x , \bm y 
 - \bm x, \alpha) + \chi \, \Phi_{pol} (\bm z - \bm x , \bm y - \bm x, \alpha) 
 \; .
\end{equation}
Substituting the wave function \peq{eq-coord wavefn} into \eq{eq-wavefn
  squared} and performing the sums gives the unpolarized part as
\begin{eqnarray}
 \label{eq-unpolarized wavefn}
 \Phi_{unp} &=& \frac{2 \, \alpha_s}{\pi} \, {\tilde m}^2 \, \bigg[ (1+\alpha^2) \, \frac{(\bm z -   \bm x) \cdot (\bm y - \bm x)}{|\bm z - \bm x| \;       
 |\bm  y - \bm x|} \, K_1 (\tilde m \, |\bm z - \bm x|) \, K_1 (\tilde m \, 
 |\bm y - \bm x|) \\ \nonumber
 && + \, (1-\alpha)^2 \, K_0 (\tilde m \, |\bm z - \bm x|) \, K_0 (\tilde m \,  
 |\bm y - \bm x|) \bigg]
\end{eqnarray}
and the transversely-polarized part as
\begin{eqnarray}
 \label{eq-polarized wavefn}
 \Phi_{pol} = \frac{2 \, \alpha_s}{\pi} \, {\tilde m}^2 \, \alpha \, (1-\alpha) \, \bigg[
 \frac{z^{2} - x^{2}}{|\bm z - \bm x|} \, K_0 (\tilde m \, |\bm y - 
 \bm x|) \, K_1 (\tilde m \, |\bm z - \bm x|) 
  + \, \frac{y^{2} - x^{2}}{|\bm y - \bm x|} \, K_1 (\tilde m \, |\bm y -
 \bm x|) \, K_0 (\tilde m \, |\bm z - \bm x|) \bigg] .
\end{eqnarray}
Note that $\Phi_{unp}$ is a scalar under rotations in the transverse
plane, whereas $\Phi_{pol}$ has an explicitly preferred
$x^{2}$-direction (i.e., the y-axis) since it ``knows'' about the
transverse polarization of the incoming quark. The $x^{2}$-axis can be
written as the direction of the ${\vec p} \times {\vec S}$ vector,
since the incoming quark with momentum $\vec p$ is moving along the
$z$-axis, while being polarized along the $x = x^1$-axis, such that
${\vec S} \, \| \, {\hat x}_1$.  We show that the unpolarized part of
the wave function squared $\Phi_{unp}$ contributes to the unpolarized
quark production cross section $d\sigma_{unp}$, while the
polarization-dependent part of the wave function squared $\Phi_{pol}$
generates the spin-asymmetric cross section $d (\Delta \sigma)$.



\subsection{Spin, Asymmetry, and C-Parity in Quark Production}
 
\label{subsec-Quark STSA (gen)}

\begin{figure}[h]
 \includegraphics[width=8cm]{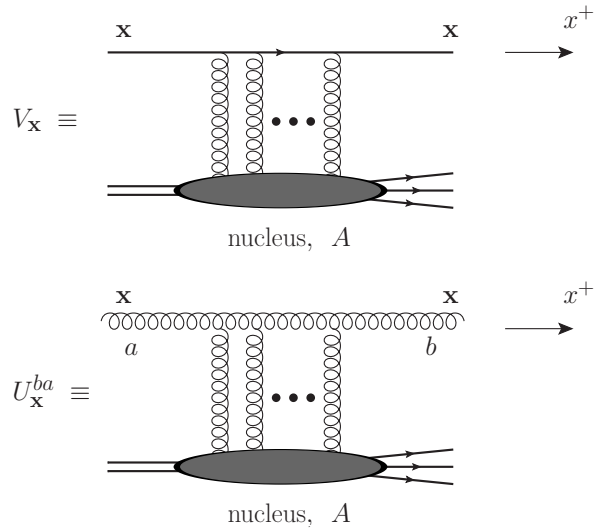}
\caption{Wilson lines resumming scattering in the small-$x$ field of the 
  target.  The quark propagator is in the fundamental representation
  (top), and the gluon propagator is in the adjoint representation
  (bottom).}
\label{fig-Wilson lines} 
\end{figure}

Having computed the $q \to q \, G$ light-cone wave function, we can
now construct the scattering cross section by allowing the wave
function to interact with the small-$x$ field of the target nucleus.
It is well known \cite{Balitsky:1996ub,Jalilian-Marian:1998cb} that
eikonal quark and gluon propagators in the background color field
$A^{\mu \, a}$ can be correspondingly written as fundamental and
adjoint path-ordered Wilson lines
\begin{subequations}
 \label{eq-Wilson lines}
 \begin{eqnarray}
   V_{\bm x} &\equiv& {\mathcal P} \exp \left[ \frac{i \, g}{2} \, \int\limits_{-\infty}^{+\infty} d x^+ \, T^a \, A^{- \, a} 
     (x^+, x^- =0, \bm x ) \right] \\
   U_{\bm x}^{ba} &\equiv& {\mathcal P} \exp \left[ \frac{i \, g}{2} \, \int\limits_{-\infty}^{+\infty} d x^+ \, t^c \, A^{- \,
       c} (x^+, x^-=0, \bm x) \right]^{ba} \, ,
 \end{eqnarray}
\end{subequations}
where $t^a$'s are the SU($N_c$) generators in the adjoint
representation and the projectile is moving along the light-cone
$x^+$-axis. In essence, this means that the the projectile's
transverse position is not altered during the scattering, and the
effect of the target field is to perform a net SU($N_c$) color
rotation on the projectile.  The Wilson lines resum these interactions
and give the total phase of that color rotation.  They are illustrated
in \fig{fig-Wilson lines}. Note that the adjoint Wilson line $U_{\bm
  x}^{ba}$ is real-valued.

The Wilson-line approach is quite generic: if the target gluon field
is quasi-classical, as in the case of the McLerran--Venugopalan (MV)
model \cite{McLerran:1993ni,McLerran:1993ka,McLerran:1994vd}, then
correlators of the Wilson lines resum powers of $\as^2 \, A^{1/3}$
corresponding to the Glauber--Mueller multiple-rescattering
approximation \cite{Mueller:1989st}. Non-linear small-$x$ evolution
resumming powers of $\as \, Y \sim \as \, \ln s$ can be included into
the correlators of the Wilson lines through the Balitsky--Kovchegov
(BK)
\cite{Balitsky:1996ub,Balitsky:1997mk,Balitsky:1998ya,Kovchegov:1999yj,Kovchegov:1999ua}
and Jalilian-Marian--Iancu--McLerran--Weigert--Leonidov--Kovner
(JIMWLK)
\cite{Jalilian-Marian:1997xn,Jalilian-Marian:1997jx,Jalilian-Marian:1997gr,Jalilian-Marian:1997dw,Jalilian-Marian:1998cb,Kovner:2000pt,Weigert:2000gi,Iancu:2000hn,Ferreiro:2001qy}
evolution equations. Thus expressing the interaction with the target
in terms of the Wilson lines \peq{eq-Wilson lines} allows for several
different levels of approximation for this interaction.

\begin{figure}[th]
 \includegraphics[width=\textwidth]{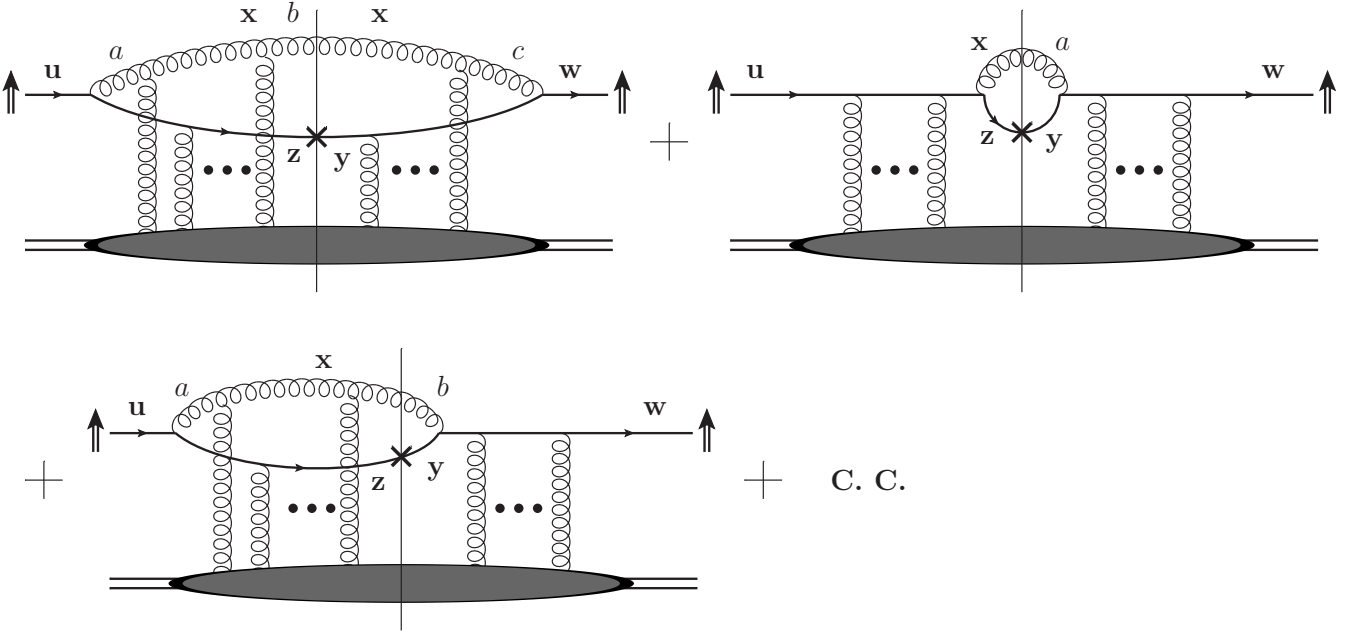}
 \caption{The cross section for quark production in the polarized quark--nucleus scattering.}
\label{fig-amplitude squared} 
\end{figure}

The scattering amplitude for quark production is composed of two
sub-processes: the splitting of \eq{eq-coord wavefn} calculated in
Sec.~\ref{subsec-Wavefn} and the Wilson line scattering
\eqref{eq-Wilson lines} of the quark and the gluon in the field of the
target.  These elements give two distinct diagrams contributing to the
scattering amplitude shown in \fig{qtoqGampl} above. To find the quark
production cross section we need to square the diagrams in
\fig{qtoqGampl}, keeping the transverse momentum of the quark fixed,
as depicted in \fig{fig-amplitude squared}. As discussed above, this
implies that the transverse coordinates of the quark are different on
both sides of the cut.  Just like in other similar calculations
\cite{Kovchegov:1998bi,Albacete:2006vv} the $q \to q \, G$ splitting
may occur either before or after the interaction with the target, both
in the amplitude and in the complex conjugate amplitude, resulting in
the four different terms shown in \fig{fig-amplitude squared}.

Using Fig.~\ref{fig-amplitude squared} we can write down the
expression for the color-averaged amplitude squared $\left\langle M^2
\right\rangle$ in terms of Wilson lines and the wave function
responsible for the splitting, remembering to reverse the sign in the
wave function for splitting occurring after the interaction:
\begin{eqnarray}
 \label{eq-amplitude squared}
 \left\langle M^2 \right\rangle &=& \frac{1}{N_c} \sum_{\lambda, \chi'} 
 \bigg[ 
 \mathrm{Tr} \left[ V_{\bm
     z} \, \psi_{\lambda \chi \chi'}^a \, \psi_{\lambda \chi \chi'}^{c \, \dagger} \, V_{\bm
     y}^\dagger \right] \, U_{\bm x}^{b a} \, U_{\bm x}^{b c} + 
 \mathrm{Tr} \left[ \psi^a_{\lambda
     \chi \chi'} \, V_{\bm u} \, V^\dagger_{\bm w} \, \psi^{a \, \dagger}_{\lambda \chi 
     \chi'} \right]
 \\ \nonumber &-&
 \mathrm{Tr} \left[ V_{\bm z} \, \psi^a_{\lambda \chi \chi'} \, V^\dagger_{\bm w} \,
   \psi^{b \, \dagger}_{\lambda \chi \chi'} \right] \, U^{b a}_{\bm x} - \mathrm{Tr} \left[
   \psi^a_{\lambda \chi \chi'} \, V_{\bm u} \, \psi^{b \, \dagger}_{\lambda \chi \chi'} \, 
   V^\dagger_{\bm y} \right] \,  U^{a b}_{\bm x} 
 \bigg] \; ,
\end{eqnarray}
where $N_c$ is the number of colors, the traces are taken over the
fundamental representation indices, and summation is implied over
repeated adjoint color indices.  Substituting Eq.~\eqref{eq-Psi vs
  Phi} into \eq{eq-amplitude squared} and using the identities
\begin{equation}
 \label{eq-identities}
 U^{b a}_{\bm x} \, T^a = V^\dagger_{\bm x} \, T^b \, V_{\bm x}
 \;\;\;\; , \;\;\;\;
 \mathrm{Tr} \left[ A \, T^a \, B \, T^a \right] = \frac{1}{2} \, 
\mathrm{Tr} A \ \mathrm{Tr} B - \frac{1}{2 N_c} \, \mathrm{Tr} \left[ A \, B \right]
\end{equation}
for arbitrary $N_c \times N_c$ matrices $A, \, B$, we find
\begin{equation}\label{eq-amplitude squared 1}
\left\langle M^2 \right\rangle = C_F \, \delta^2 \big[ \bm u - \bm x -
 \alpha \, (\bm z - \bm x) \big] \, \delta^2 \big[ \bm w - \bm x -
 \alpha \, (\bm y - \bm x) \big] \, \Phi_\chi (\bm z - \bm x , \bm y - \bm x) 
 \  \mathcal{I}^{(q)}
\end{equation}
where the factor responsible for the quark's interaction with the
target, denoted by $\mathcal{I}^{(q)}$, is given by
\begin{eqnarray}
\mathcal{I}^{(q)} =  \left\langle \frac{1}{N_c} \, \mathrm{Tr} \, 
\left[ V_{\bm z} \, V^\dagger_{\bm y} \right] 
+ \frac{1}{N_c} \, \mathrm{Tr} \, \left[ V_{\bm u} \, V^\dagger_{\bm w} \right]  
- \frac{1}{2 \, N_c \, C_F} \,  \mathrm{Tr} \, \left[ V_{\bm z} \, V^\dagger_{\bm x} \right]  
\, \mathrm{Tr} \, \left[ V_{\bm x} \, V^\dagger_{\bm w} \right] + 
\frac{1}{2 \, N_c^2 \, C_F} \, 
\mathrm{Tr} \, \left[ V_{\bm z} \, V^\dagger_{\bm w} \right] \right. \notag \\
\left. - \frac{1}{2 \, N_c \, C_F} \,  
\mathrm{Tr} \, \left[ V_{\bm u} \, V^\dagger_{\bm x} \right]  \, 
\mathrm{Tr} \, \left[ V_{\bm x} \, V^\dagger_{\bm y} \right] + 
\frac{1}{2 \, N_c^2 \, C_F} \, 
\mathrm{Tr} \, \left[ V_{\bm u} \, V^\dagger_{\bm y} \right] \right\rangle .
\end{eqnarray}
Here $C_F = (N_c^2 - 1)/2 N_c$ is the fundamental Casimir operator of
SU($N_c$), and the angle brackets on the right denote averaging over
the field configurations of the target.

Defining the $S$-matrix operator for a fundamental-representation
color dipole by
\begin{equation}
  \label{Ddef}
  {\hat D}_{\bm x \, \bm y} \equiv \frac{1}{N_c} \, \mathrm{Tr} \, \left[ V_{\bm x}
 \, V^\dagger_{\bm y} \right]
\end{equation}
we can rewrite $\mathcal{I}^{(q)}$ more compactly as
\begin{equation}
  \label{Iq}
  \mathcal{I}^{(q)} = \left\langle {\hat D}_{\bm z \, \bm y} + {\hat D}_{\bm u \, \bm w} 
- \frac{N_c}{2 \, C_F} \, {\hat D}_{\bm z \, \bm x} \, {\hat D}_{\bm x \, \bm w} + 
 \frac{1}{2 \, N_c \, C_F} \, {\hat D}_{\bm z \, \bm w} - 
\frac{N_c}{2 \, C_F} \, {\hat D}_{\bm u \, \bm x} \, {\hat D}_{\bm x \, \bm y} + 
 \frac{1}{2 \, N_c \, C_F} \, {\hat D}_{\bm u \, \bm y} \right\rangle . 
\end{equation}
As we have already mentioned, this interaction with the target can be
evaluated either in the Glauber--Mueller multiple-rescattering
approximation or using the JIMWLK evolution equation.

The expression \peq{Iq} simplifies in 't Hooft's large-$N_c$ limit, in
which the correlators of several single-trace operators factorize,
such that, for instance, $\langle {\hat D}_{\bm u \, \bm x} \, {\hat
  D}_{\bm x \, \bm y} \rangle = \langle {\hat D}_{\bm u \, \bm x}
\rangle \, \langle {\hat D}_{\bm x \, \bm y} \rangle$
\cite{Balitsky:1996ub,Kovchegov:1999yj,Weigert:2005us}.  Defining
\begin{equation}
 \label{eq-dipole propagator}
 D_{\bm x \, \bm y} \equiv \left\langle {\hat D}_{\bm x \, \bm y} \right\rangle
= \frac{1}{N_c} \left\langle \mathrm{Tr} \, \left[ V_{\bm x}
 \, V^\dagger_{\bm y} \right] \right\rangle
\end{equation}
we rewrite \eq{Iq} in the large-$N_c$ limit as
\begin{equation}
 \label{eq-interaction}
 \mathcal{I}^{(q)} \bigg|_{\mbox{large}-N_c} = D_{\bm z \, \bm y} + D_{\bm u \, \bm w} -
 D_{\bm z \, \bm x} \, D_{\bm x \, \bm w} - D_{\bm u \,
 \bm x} \, D_{\bm x \, \bm y} \; .
\end{equation}

To compute the quark production cross sections, we need to Fourier
transform the coordinate space amplitude squared of \eq{eq-amplitude
  squared 1} back to momentum space and include the appropriate
kinematic factors. This is accomplished by
\cite{Jalilian-Marian:2005jf} 
\begin{equation}
 \label{eq-inverse Fourier transform}
 \frac{d\sigma^{(q)}}{d^2 k \, d y_q} 
 = \frac{1}{2 \, (2\pi)^3} \, \frac{\alpha}{1-\alpha} \, 
 \int d^2 x \, d^2 y \, d^2 z \, d^2 u \, d^2 w \, e^{-i \bm k \cdot (\bm z - 
 \bm y)} \, e^{i \bm p \cdot (\bm u - \bm w)} \, 
 \langle M^2 \rangle \; 
\end{equation}
with $\bm k$ and $y_q$ the transverse momentum and rapidity of the
produced quark. Integrating over the delta functions in
\eqref{eq-amplitude squared 1} imposes the kinematic constraints
\begin{subequations}
 \label{eq-kinematic constraints}
 \begin{eqnarray}
  \bm u = \bm x + \alpha \, (\bm z - \bm x) \\ 
  \bm w = \bm x + \alpha \, (\bm y - \bm x)
 \end{eqnarray}
\end{subequations}
which relate the quark coordinates before and after the $q \to q \, G$
splitting and describe the non-eikonal quark recoil.  To make the
incoming quark transversely polarized we need to put the transverse
momentum of the incoming quark to zero: $\bm p = \bm 0$. We thus
obtain the general result for the quark production in the $q^\uparrow
+ A$ scattering
\begin{equation}
 \label{eq-cross section}
 \frac{d\sigma^{(q)}}{d^2 k \, d y_q} = \frac{C_F}{2 \, (2\pi)^3} \, 
 \frac{\alpha}{1-\alpha} \, \int
 d^2 x \, d^2 y \, d^2 z \, e^{-i \bm k \cdot (\bm z - \bm y)} \,
 \Phi_\chi (\bm z - \bm x \, , \, \bm y - \bm x, \alpha) \
 \mathcal{I}^{(q)} (\bm x \, , \, \bm y \, , \, \bm z)
\end{equation}
with $\Phi_\chi$ from \eq{eq-wavefn squared} and $\mathcal{I}^{(q)}$
from \eq{Iq}. The expression \peq{eq-cross section} contains multiple
rescatterings and non-linear small-$x$ evolution between the
projectile and the target. Note that it does not resum the small-$x$
evolution between the produced quark and the projectile (which can be
included following \cite{Albacete:2006vv}), and hence is not valid for
very small $\alpha$ (i.e., the values of $\alpha$ are restricted by
$\as \, \ln 1/\alpha \ll 1$). Since, as we will see below, both the
experimental STSA and STSA resulting from our production mechanism
fall off with decreasing $\alpha$, the region of interest in this work
corresponds to $\alpha$ not being very small, where \eq{eq-cross
  section} is fully applicable.

Now we are in a position to analyze the symmetry properties of the
wave function and the interaction.  There are two relevant symmetries
to consider: $\bm k \rightarrow - \bm k$ or ``$k_T$-parity'' and $\chi
\rightarrow - \chi$ spin-flip.  Under $k_T$-parity, the quark and
antiquark coordinates $\bm z$ and $\bm y$ get interchanged, $\bm z
\leftrightarrow \bm y$. (Note that, due to Eqs.~\peq{eq-kinematic
  constraints}, this also implies that $\bm u \leftrightarrow \bm w$.)
From our previous calculation of the wave function
\eqref{eq-unpolarized wavefn}, \eqref{eq-polarized wavefn}, we note
that $\Phi_{unp}$ is spin-independent, a scalar under rotations in the
transverse plane, and $k_T$-even.  Similarly, $\chi \, \Phi_{pol}$ is
odd under spin-flip, a vector under transverse rotations, and
$k_T$-even.  We can explicitly (anti-)symmetrize the interaction with
the target under $\bm z \leftrightarrow \bm y$ interchange
\begin{equation}
 \label{eq-defn interaction parity}
 \mathcal{I}_{symm \, / \, anti} \equiv \frac{1}{2} \bigg( \mathcal{I} \pm (\bm z 
 \leftrightarrow \bm y) \bigg). 
\end{equation}
To do so we first need to decompose each dipole $S$-matrix into the
even and odd pieces under the exchange of transverse coordinates,
which corresponds to the $C$-parity operation exchanging the quark and
the anti-quark \cite{Kovchegov:2003dm,Hatta:2005as}:
\begin{subequations}
 \label{eq-Pomeron and Odderon}
 \begin{eqnarray}
  {\hat D}_{\bm x \, \bm y} &\equiv& {\hat S}_{\bm x \, \bm y} + i \,
  {\hat O}_{\bm x \, \bm y} \\
  {\hat S}_{\bm x \, \bm y} &\equiv& \frac{1}{2} \, ( {\hat D}_{\bm x \, \bm y} +
  {\hat D}_{\bm y \, \bm x}) \label{Sdef} \\
  {\hat O}_{\bm x \, \bm y} &\equiv& \frac{1}{2i} \, ( {\hat D}_{\bm x \, \bm y} -
  {\hat D}_{\bm y \, \bm x}) \; . \label{Odef}
 \end{eqnarray}
\end{subequations}
The $C$-even real part of the target field-averaged $S$-matrix $S_{\bm
  x \, \bm y} \equiv \langle {\hat S}_{\bm x \, \bm y} \rangle$ is
responsible for the total unpolarized cross section of the
dipole--target interactions. Its small-$x$ evolution is given by the
BK/JIMWLK equations.  The $C$-odd imaginary part of the
target-averaged $S$-matrix $O_{\bm x \, \bm y} \equiv \langle {\hat
  O}_{\bm x \, \bm y} \rangle$ is known as the odderon interaction
\cite{Lukaszuk:1973nt,Nicolescu:1990ii,Ewerz:2003xi}. The small-$x$
evolution equation for $O_{\bm x \, \bm y}$ was constructed in
\cite{Kovchegov:2003dm,Hatta:2005as,Kovner:2005qj}, and, in the linear
approximation, was found to be identical to the dipole BFKL equation
\cite{Mueller:1994rr} with $C$-odd initial conditions. The intercept
of the linearized odderon evolution was found to be $\alpha_O - 1 =0$,
in agreement with the solution of the
Bartels--Kwiecinski--Praszalowicz (BKP)
\cite{Bartels:1978fc,Kwiecinski:1980wb} equation for the odderon found
in \cite{Bartels:1999yt}. For the current status of the experimental
searches for the QCD odderon and for an overview of the theory see
\cite{Ewerz:2003xi}.

With these explicitly symmetrized elements, it is straightforward to
construct the symmetric and antisymmetric parts of the interaction
with the target \peq{Iq} for quark production:
\begin{subequations}
 \label{eq-symmetrized interaction}
 \begin{eqnarray}
  \mathcal{I}_{symm}^{(q)} &=& \left\langle {\hat S}_{\bm z \, \bm y} +
  {\hat S}_{\bm u \, \bm w} - \frac{N_c}{2 \, C_F} \, \left( {\hat S}_{\bm z \, \bm x} \, 
  {\hat S}_{\bm x \, \bm w} - {\hat O}_{\bm z \,  \bm x} \, {\hat O}_{\bm x \, \bm w} \right) 
  + \frac{1}{2 \, N_c \, C_F} \, {\hat S}_{\bm z \, \bm w} \right. \notag \\ 
  && \left. - \frac{N_c}{2 \, C_F} \, \left(
  {\hat S}_{\bm u \, \bm x} \, {\hat S}_{\bm x \, \bm y} -  {\hat O}_{\bm u \, \bm x} \, 
  {\hat O}_{\bm x \, \bm y} \right) 
  + \frac{1}{2 \, N_c \, C_F} \, {\hat S}_{\bm u \, \bm y} \right\rangle , \\
  \mathcal{I}_{anti}^{(q)} &=& i \, \left\langle {\hat O}_{\bm z \, \bm y} +
  {\hat O}_{\bm u \, \bm w} - \frac{N_c}{2 \, C_F} \, 
  \left( {\hat O}_{\bm z \, \bm x} \, {\hat S}_{\bm x \, \bm w} 
   + {\hat S}_{\bm z \, \bm x} \, {\hat O}_{\bm x \, \bm w} \right) 
  + \frac{1}{2 \, N_c \, C_F} \, {\hat O}_{\bm z \, \bm w} \right. \notag \\ 
  && \left. - \frac{N_c}{2 \, C_F} \, 
  \left(
  {\hat O}_{\bm u \, \bm x} \, {\hat S}_{\bm x \, \bm y} +  {\hat S}_{\bm u \, \bm x} \,
  {\hat O}_{\bm x \, \bm y} \right) 
  + \frac{1}{2 \, N_c \, C_F} \, {\hat O}_{\bm u \, \bm y} \right\rangle \; .
 \end{eqnarray}
\end{subequations}
In the large-$N_c$ limit these expressions simplify to
\begin{subequations}
 \label{eq-symmetrized interaction-N}
 \begin{eqnarray}
  \mathcal{I}_{symm}^{(q)}\bigg|_{\mbox{large}-N_c} &=& S_{\bm z \, \bm y} +
  S_{\bm u \, \bm w} - S_{\bm z \, \bm x} \, S_{\bm x \, \bm w} -
  S_{\bm u \, \bm x} \, S_{\bm x \, \bm y} + O_{\bm z \,
  \bm x} \, O_{\bm x \, \bm w} + O_{\bm u \, \bm x} \, 
  O_{\bm x \, \bm y} , \label{qsymm} \\
  \mathcal{I}_{anti}^{(q)} \bigg|_{\mbox{large}-N_c} &=& i \left[ O_{\bm z \, \bm y} +
  O_{\bm
  u \, \bm w} - O_{\bm z \, \bm x} \, S_{\bm x \, \bm w} -
  O_{\bm u \, \bm x} \, S_{\bm x \, \bm y} - S_{\bm z \,
  \bm x} \, O_{\bm x \, \bm w} - S_{\bm u \, \bm x} \,
  O_{\bm x \, \bm y} \right] \; . \label{qanti}
 \end{eqnarray}
\end{subequations}

Because the production cross section preserves azimuthal rotational
symmetry, any physical observable must behave the same way under both
$k_T$-parity and spin flip; that is, the only terms which are non-zero
are those which are \emph{even} under the combined operation $k_T \,
\otimes$~(spin flip). Note that $\Phi_{unp}$ is even under the
spin-flip, while $\chi \, \Phi_{pol}$ is odd under spin-flip, and both
are even under $k_T$-parity. Hence in the $\Phi_\chi \,
\mathcal{I}^{(q)}$ product in \eq{eq-cross section} the terms which
give non-zero contributions to the cross section are $\Phi_{unp} \,
\mathcal{I}^{(q)}_{symm}$ and $\chi \, \Phi_{pol} \,
\mathcal{I}^{(q)}_{anti}$.  The other terms $\Phi_{unp} \,
\mathcal{I}^{(q)}_{anti}$ and $\chi \, \Phi_{pol} \,
\mathcal{I}^{(q)}_{symm}$ contain a contradiction between their vector
structure and their $k_T$-parity, so they vanish identically: for
instance, the contribution to the cross section coming from
$\Phi_{unp} \, \mathcal{I}^{(q)}_{anti}$ is odd under $k_T$-parity and
is a function of $k_T^2$ only, which is possible only if it is zero.
Similarly, the contribution of $\chi \, \Phi_{pol} \,
\mathcal{I}^{(q)}_{symm}$ is even under $k_T$-parity and is a vector
in transverse space, which implies that it is also zero. Of the
nonzero terms, $\Phi_{unp} \, \mathcal{I}^{(q)}_{symm}$ does not
change its sign under spin-flip; it generates the symmetric part of
the distribution $d\sigma_{unp}$ in \eq{eq-Defn STSA}. At the same
time $\chi \, \Phi_{pol} \, \mathcal{I}^{(q)}_{anti}$ does change sign
under spin-flip; it generates the transverse spin asymmetry we are
looking for.

Knowing these symmetry properties, we can explicitly construct the
spin-dependent and spin-averaged cross sections $d(\Delta \sigma)$ and
$d\sigma_{unp}$ for quark production from their definitions
\eqref{eq-Defn STSA} obtaining
\begin{subequations}
 \label{eq-observables}
 \begin{eqnarray}
  d(\Delta\sigma^{(q)}) &=& \frac{C_F}{(2\pi)^3} \, \frac{\alpha}{1-\alpha} \, 
  \int d^2 x \, d^2 y \, d^2 z
  \, e^{-i \bm k \cdot (\bm z - \bm y)} \, \Phi_{pol} (\bm z - \bm x \, , \, \bm y - \bm x, \alpha) \
  \mathcal{I}^{(q)}_{anti} (\bm x \, , \, \bm y \, , \, \bm z) \label{dsigmaq} \\
  d\sigma^{(q)}_{unp} &=& \frac{C_F}{2 \, (2\pi)^3} \, \frac{\alpha}{1-\alpha} \, 
  \int d^2 x \, d^2 y \, d^2 z
  \, e^{-i \bm k \cdot (\bm z - \bm y)} \, \Phi_{unp} (\bm z - \bm x \, , \, \bm y - \bm x, \alpha) \
  \mathcal{I}^{(q)}_{symm} (\bm x \, , \, \bm y \, , \, \bm z) \; , \label{dsigmaq_unp}
 \end{eqnarray}
\end{subequations}
where the wave functions squared are given by Eqs.
\eqref{eq-unpolarized wavefn}, \eqref{eq-polarized wavefn}, and the
interactions are given by Eqs.~\eqref{eq-symmetrized interaction} (and
by Eqs.~\peq{eq-symmetrized interaction-N} in the large-$N_c$ limit).
Eqs.~\eqref{eq-observables} are one of the main results of this work:
together with \eq{eq-Defn STSA} they give the single-transverse spin
asymmetry $A_N$ generated in quark production by the $C$-odd CGC
interactions with the target.

The mechanism for the generation of the STSA in
Eqs.~\eqref{eq-observables} is different from both the well-known
Sivers \cite{Sivers:1989cc,Sivers:1990fh} and Collins
\cite{Collins:1992kk} effects. It appears difficult (if not
impossible) to absorb the interactions of \fig{fig-amplitude squared}
into the projectile wave function (distribution function): hence our
result is different from the Sivers effect. In the above calculation
the asymmetry is generated before fragmentation; hence the STSA
resulting from Eqs.~\eqref{eq-observables} cannot be due to Collins
effect either.  As we will see below, the non-zero part of
\eq{dsigmaq} stems from the multiple interactions with the target
(higher-twist effects), and its contribution is in fact zero in the
linearized (leading-twist) approximation. In this sense the above
mechanism for generating STSA is similar in spirit to the higher-twist
mechanisms of
\cite{Efremov:1981sh,Efremov:1984ip,Qiu:1991pp,Ji:1992eu,Qiu:1998ia,Brodsky:2002cx,Collins:2002kn,Koike:2011mb,Kanazawa:2000hz,Kanazawa:2000kp},
though a detailed comparison of the diagrams appears to indicate that
the two approaches are, in fact, different.

We have shown explicitly that the single-transverse spin asymmetry
$A_N$ occurs in the CGC framework as a coupling between the transverse
spin of the projectile and a $C$-odd interaction with the target,
driven by the odderon.\footnote{In the past, the relation between the
  odderon and the single and double transverse spin asymmetries was
  investigated in
  \cite{Ahmedov:1999ne,Jarvinen:2006dm,Leader:1999ua,Buttimore:1998rj,Trueman:2007fr}
  in the pomeron and reggeon formalism.} Note that to date there is no
unambiguous experimental evidence for the QCD odderon. If our
mechanism for generating STSA can be isolated experimentally from
other contributions, it may constitute the first direct observation of
the QCD odderon! To make such a distinction possible, one needs to
determine phenomenological characteristics of our mechanism, such as
its rapidity, energy, and centrality dependence; some of this work
will be carried out below, while the rest, along with a proper
phenomenological implementation of our results, is left for future
work.


Within LCPT, the real-virtual cancellations or ``crossing symmetry''
embodied in Eq. \eqref{eq-identities} allow particles appearing in the
complex conjugate amplitude $M^*$ to be rewritten as their
charge-conjugate particles appearing in the amplitude $M$.  This is
what gives rise to the natural dipole degrees of freedom within CGC,
and in terms of the asymmetry, this feature translates the
$k_T$-parity of the cross section into the $C$-parity of the
interaction with the target. This is the reason the odderon appears
naturally in the expression for the asymmetry.

In the literature it is often emphasized that the STSA $A_N$ is odd
under the time reversal transformation $T$ \cite{Collins:1992kk}. To
elucidate how the obtained result transforms under $T$ we note that
$T$ reverses the directions of ${\vec S}$, ${\vec p}$, and ${\vec k}$:
\begin{align}
  \label{eq:T}
  {\vec S} \, {\overset{\scriptscriptstyle T} \rightarrow} \, - {\vec
    S}, \ \ \ {\vec p} \, {\overset{\scriptscriptstyle T} \rightarrow}
  \, - {\vec p}, \ \ \ {\vec k} \, {\overset{\scriptscriptstyle T}
    \rightarrow} \, - {\vec k}.
\end{align}
Since $({\vec p} \times {\vec S}) \parallel {\hat x}_2$ is invariant
under $T$, the only effect of time-reversal on the spin-dependent
cross section \eqref{dsigmaq} is
\begin{align}
  \label{eq:Tsigma}
  d(\Delta\sigma^{(q)}) ({\bm k}) \ \ {\overset{\scriptscriptstyle T}
    \rightarrow} \ \ d(\Delta\sigma^{(q)}) (- {\bm k}),
\end{align}
since it is a real Lorentz-scalar momentum-space quantity.  Therefore
time reversal is equivalent to the $k_T$-parity transformation
discussed above, and, hence, to the ${\bm z} \leftrightarrow {\bm y}$
interchange. Because of the odderon exchange, the interaction with the
target $\mathcal{I}^{(q)}_{anti} (\bm x \, , \, \bm y \, , \, \bm z)$
is ${\bm z} \leftrightarrow {\bm y}$ anti-symmetric, and is,
therefore, $T$-odd. This results in the spin-dependent cross section
being $T$-odd too,
\begin{align}
  \label{eq:Tsigma2}
  d(\Delta\sigma^{(q)}) ({\bm k}) \ \ {\overset{\scriptscriptstyle T}
    \rightarrow} \ \ d(\Delta\sigma^{(q)}) (- {\bm k}) = -
  d(\Delta\sigma^{(q)}) ({\bm k}),
\end{align}
leading to the $T$-odd STSA $A_N$, in agreement with the standard
expectations \cite{Collins:1992kk}. 

It is interesting to note that in the high-energy approximation
considered here the application of time reversal to dipole correlators
is equivalent to the application of $C$-parity, such that the STSA
arises from the odderon exchange, which is both $T$- and $C$-odd. As
one can check explicitly, under time reversal Wilson lines transform
as
\begin{align}
  \label{eq:WT}
  V_{\bm x} \ \ {\overset{\scriptscriptstyle T} \rightarrow} \ \ T \,
  V_{\bm x} \, T^{-1} = (V_{\bm x}^\dagger)^* = V_{\bm x}^T
\end{align}
with $T$ in the superscript denoting transposition. Note that if the
original Wilson line was a future-pointing integral along the
$x^+$-axis, the time-reversed Wilson line on the right of \eq{eq:WT}
can be thought of as a complex conjugate of a past-pointing Wilson
line or a transpose of the future-pointing Wilson line. The
integration in either interpretation runs along the $x^-$ axis: we
will not show or refer to such $x^+ \leftrightarrow x^-$ interchange
explicitly, as it can be eliminated by a simple relabeling of the
$x^3$-axis. Since the averaging over the nuclear target is $T$-even,
which can be inferred from \cite{McLerran:1993ka,McLerran:1993ni}, the
(coordinate-space) time reversal of the odderon amplitude $O_{\bm x \,
  \bm y} = \langle {\hat O}_{\bm x \, \bm y} \rangle$ applies only to
the odderon operator, such that
\begin{align}
  \label{eq:OT}
  \langle {\hat O}_{\bm x \, \bm y} \rangle \ \
  {\overset{\scriptscriptstyle T} \rightarrow} \ \ - \langle {\hat
    O}_{\bm x \, \bm y} \rangle,
\end{align}
because $\langle {\hat D}_{\bm x \, \bm y} \rangle$ is invariant under
transposition and $T$ is antilinear ($i \ {\overset{\scriptscriptstyle
    T} \rightarrow} \ -i$). Thus we see that the $T$-odd odderon
exchange leads to the $T$-odd STSA observable $A_N$.

Finally, the reader may wonder whether the cross section in
\eq{dsigmaq} is in fact non-zero. While it is very difficult to carry
out the integration in \eq{dsigmaq} exactly, we instead will evaluate
\eq{dsigmaq} approximately in Sec.~\ref{sec-Estimates}, showing that
the cross section and the corresponding STSA $A_N$ are in fact
non-zero. However, first we would like to derive the analogues of
Eqs.~\eqref{eq-observables} for gluon and prompt photon production.



\subsection{STSA in Direct Quark Production}
\label{Sec_direct}

\begin{figure}[th]
 \includegraphics[width=.7 \textwidth]{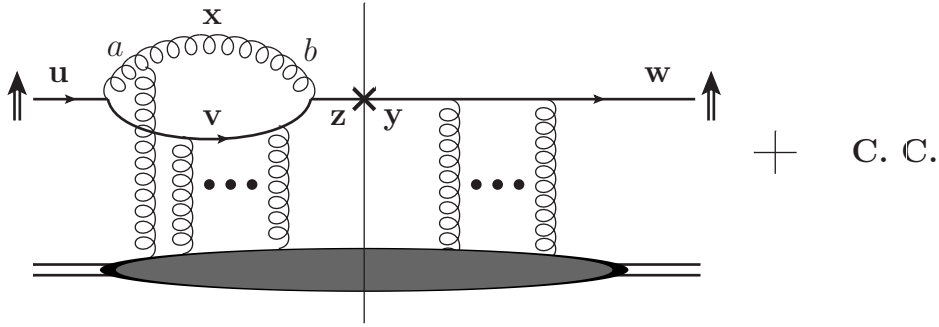}
 \caption{Direct quark production in the polarized quark--nucleus scattering.}
\label{direct_quark} 
\end{figure}

Another channel for quark production is shown in \fig{direct_quark}
and contains a ``virtual'' gluon correction, without the gluon being
present in the final state. The gluon and quark in the loop may still
interact with the target, as depicted in \fig{direct_quark}. It can be
shown that similar diagrams with the gluon loop located either
completely before or after the interaction with the target do not
contribute to the STSA and are, therefore, not considered here. The
process illustrated in \fig{direct_quark} only leads to quark
production at $\alpha =1$, and may seem to be negligible if, in order
to avoid the projectile's fragmentation region, we constrain ourselves
to $\alpha < 1$ kinematics for the produced quark.  However,
remembering that for phenomenological applications one would have to
convolute our quark production cross section \peq{eq-observables} with
the quark distribution in the projectile proton, we see that $\alpha
=1$ production may still lead to non-negligible quark production at
Feynman-$x$ less than one, possibly avoiding mixing with the proton's
fragmentation region.

The calculation of the cross section for the process in
\fig{direct_quark} can be straightforwardly carried out along the
lines specified above, yielding for ${\bm p} =0$
\begin{align}
  \label{direct1}
  \frac{d\sigma^{(q) \, direct}}{d^2 k \, d y_q} = \frac{C_F}{(2 \,
    \pi)^2} \, \delta (1 - \alpha) \, \int d^2 x \, d^2 y \, d^2 z \,
  d^2 u \, d^2 v \, d^2 w \, e^{-i \bm k \cdot (\bm z - \bm y)} \,
  \int\limits_0^1 \frac{d \alpha'}{4 \, \pi \, (1-\alpha')} \,
  \Phi_\chi ({\bm v} - {\bm x}, {\bm v} - {\bm x}, \alpha') \notag \\
  \times \, \bigg[ \delta^2 [(\bm u - \bm x) - \alpha' \, (\bm v - \bm
  x)] \, \delta^2 [(\bm z - \bm x) - \alpha' \, (\bm v - \bm x)] \,
  \delta^2 ({\bm y} - {\bm w}) \, \, \mathcal{I}^{(q) \, direct}_{1}
  (\bm x \, , \, \bm y \, , \, \bm v) \notag \\ + \, \delta^2 [(\bm w
  - \bm x) - \alpha' \, (\bm v - \bm x)] \, \delta^2 [(\bm y - \bm x)
  - \alpha' \, (\bm v - \bm x)] \, \delta^2 ({\bm z} - {\bm u}) \, \,
  \mathcal{I}^{(q) \, direct}_{2} (\bm x \, , \, \bm z \, , \, \bm v)
  \bigg]
\end{align}
with $\alpha'$ the longitudinal momentum fraction of the incoming
quark carried by the quark in the loop, $y_q \sim \ln 1/\alpha$ (up to
an additive term defining the zero rapidity direction), and the
interactions with the target
\begin{subequations}
\begin{align}
  \label{direct2}
  \mathcal{I}^{(q) \, direct}_{1} (\bm x \, , \, \bm y \, , \, \bm v)
  = \left\langle \frac{N_c}{2 \, C_F} \, {\hat D}_{{\bm v} \, {\bm x}}
    \, {\hat D}_{{\bm x} \, {\bm y}} - \frac{1}{2 \, N_c \, C_F} \,
    {\hat D}_{{\bm v} \, {\bm y}} \right\rangle , \\
  \mathcal{I}^{(q) \, direct}_{2} (\bm x \, , \, \bm z \, , \, \bm v)
  = \left\langle \frac{N_c}{2 \, C_F} \, {\hat D}_{{\bm x} \, {\bm v}}
    \, {\hat D}_{{\bm z} \, {\bm x}} - \frac{1}{2 \, N_c \, C_F} \,
    {\hat D}_{{\bm z} \, {\bm v}} \right\rangle .
\end{align}
\end{subequations}

To obtain the contribution to the numerator of STSA in \eq{eq-Defn
  STSA} from \eq{direct1} we keep the spin-dependent part of the wave
function squared, $\Phi_{pol}$, and anti-symmetrize the integrand with
respect to the ${\bm z} \leftrightarrow {\bm y}$ interchange. This
gives
\begin{align}
  \label{direct3}
  d(\Delta & \, \sigma^{(q) \, direct}) = \frac{C_F}{2 \, (2 \,
    \pi)^3} \, \delta (1 - \alpha) \, \int d^2 x \, d^2 y \, d^2 z \,
  e^{-i \bm k \cdot (\bm z - \bm y)} \, \int\limits_0^1 \frac{d
    \alpha'}{(1-\alpha') \, \alpha^{\prime \, 2}} \notag \\ & \times
  \, \bigg[ \Phi_{pol} \left(\frac{{\bm z} - {\bm x}}{\alpha'},
    \frac{{\bm z} - {\bm x}}{\alpha'}, \alpha' \right) \,
  \mathcal{I}^{(q) \, direct}_{anti} ( \bm x \, , \, \bm y \, , \,
  {\bm z}) - \Phi_{pol} \left( \frac{{\bm y} - {\bm x}}{\alpha'},
    \frac{{\bm y} - {\bm x}}{\alpha'}, \alpha' \right) \,
  \mathcal{I}^{(q) \, direct}_{anti} ( \bm x \, , \, \bm z \, , \,
  {\bm y}) \bigg]
\end{align}
with the anti-symmetrized interaction given by
\begin{align}
  \label{direct4}
  \mathcal{I}^{(q) \, direct}_{anti} ( \bm x \, , \, \bm y \, , \,
  {\bm z}) \equiv \mathcal{I}^{(q) \, direct}_{1} \left( \bm x \, , \,
    \bm y \, , \, \bm x + \frac{{\bm z} - {\bm x}}{\alpha'} \right) -
  \mathcal{I}^{(q) \, direct}_{2} \left( \bm x \, , \, \bm y \, , \,
    \bm x + \frac{{\bm z} - {\bm x}}{\alpha'} \right) \notag \\ = i \,
  \left\langle \frac{N_c}{C_F} \left[ {\hat S}_{{\bm x}, \, {\bm y}}
      \, {\hat O}_{{\bm x} + \frac{1}{\alpha'} \, ({\bm z} - {\bm x})
        , \, {\bm x}} + {\hat S}_{{\bm x}, \, {\bm x} +
        \frac{1}{\alpha'} \, ({\bm z} - {\bm x})} \, {\hat O}_{{\bm x}
        , \, {\bm y}} \right] + \frac{1}{N_c \, C_F} \, {\hat O}_{{\bm
        y}, \, {\bm x} + \frac{1}{\alpha'} \, ({\bm z} - {\bm x})}
  \right\rangle .
\end{align}
Note again that the single transverse spin asymmetry is due to the
odderon-mediated interactions with the target.

\subsection{STSA in Gluon and Photon Production}
 
\label{subsec-Gluon Photon STSA (gen)}

Having laid out the methodology in Sections~\ref{subsec-Quark STSA
  (gen)} and \ref{Sec_direct}, we can now perform similar calculations
of STSA for the cases of gluon and photon production.

\begin{figure}[h]
  \includegraphics[width=\textwidth]{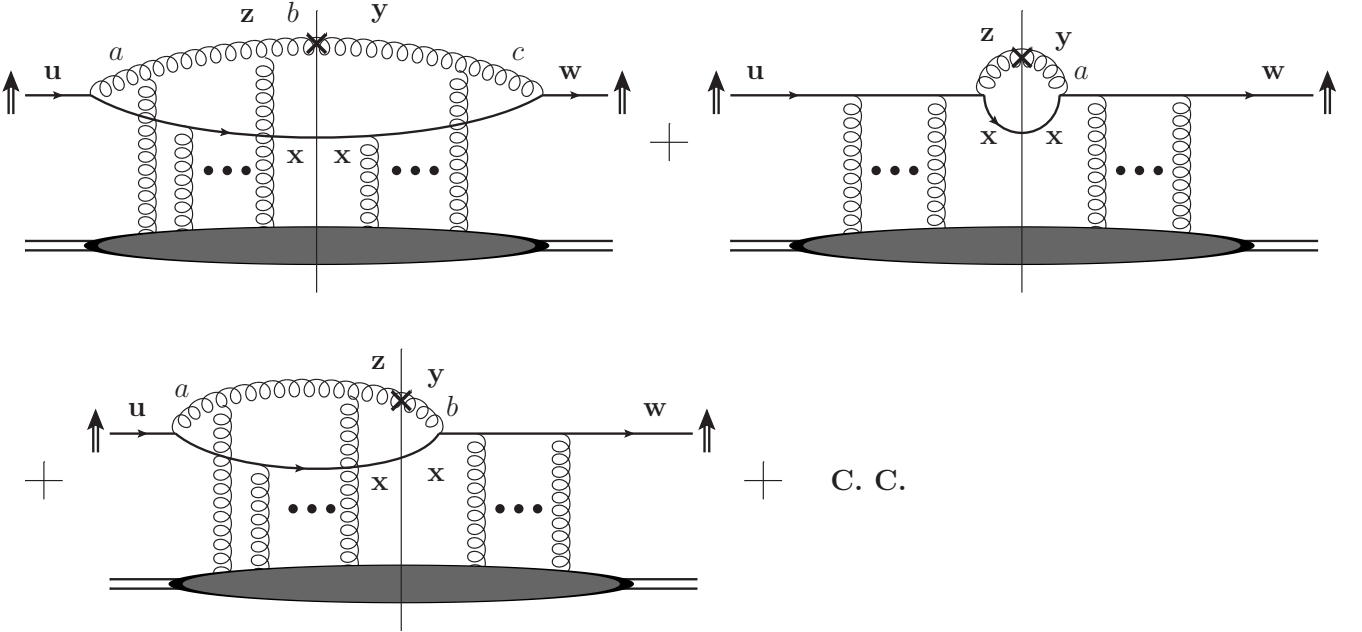}
 \caption{Diagrams contributing to the gluon / photon production cross section.}
\label{fig-gluon diagrams} 
\end{figure}

We begin with the gluon production. The gluon production diagrams are
shown in \fig{fig-gluon diagrams}. Since now we tag on the gluon, its
transverse-space positions are different on both sides of the cut, now
denoted $\bm z$ and $\bm y$, while the untagged quark has the same
transverse positions $\bm x$ in the amplitude and in the complex
conjugate amplitude. We see that to obtain the gluon production cross
section from the quark production expression found in the previous
Section, we need to interchange
\begin{equation}
  \label{interchange}
  {\bm z} \leftrightarrow {\bm x} \ \ \ \mbox{and} \ \ \  {\bm y} \leftrightarrow {\bm x}
\end{equation}
in the wave function and its complex conjugate correspondingly. In
addition, since we are interested in the differential cross section
per unit gluon rapidity $y_G$, we use
\begin{equation}
  \label{Gtoq_rap}
  d y_G = \frac{\alpha}{1 - \alpha} \, d y_q
\end{equation}
(with $\alpha$ still the fraction of the incoming quark's longitudinal
momentum carried by the final state quark).

These modifications lead to the following expressions for the
polarization-dependent and unpolarized cross sections for gluon
production
\begin{subequations}
 \label{Gprod}
 \begin{eqnarray}
  d(\Delta\sigma^{(G)}) &=& \frac{C_F}{(2\pi)^3} \,  
  \int d^2 x \, d^2 y \, d^2 z
  \, e^{-i \bm k \cdot (\bm z - \bm y)} \, \Phi_{pol} (\bm x - \bm z \, , \, \bm x - \bm y, \alpha) \
  \mathcal{I}^{(G)}_{anti} (\bm x \, , \, \bm y \, , \, \bm z) \label{dsigmaG} \\
  d\sigma^{(G)}_{unp} &=& \frac{C_F}{2 \, (2\pi)^3} \, 
  \int d^2 x \, d^2 y \, d^2 z
  \, e^{-i \bm k \cdot (\bm z - \bm y)} \, \Phi_{unp} (\bm x - \bm z \, , \, \bm x - \bm y, \alpha) \
  \mathcal{I}^{(G)}_{symm} (\bm x \, , \, \bm y \, , \, \bm z) \; , \label{Gunp}
 \end{eqnarray}
\end{subequations}
with $\Phi_{pol}$ and $\Phi_{unp}$ still given by
Eqs.~\peq{eq-polarized wavefn} and \peq{eq-unpolarized wavefn}.

The interaction with the target for the gluon production case can be
calculated along the similar lines to the above calculation of quark
production by using \fig{fig-gluon diagrams}, yielding
\begin{equation}
  \label{IG}
  \mathcal{I}^{(G)} = \left\langle {\hat D}_{\bm u \, \bm w} + 
\frac{N_c}{2 \, C_F} \, {\hat D}_{\bm z \, \bm y} \, {\hat D}_{\bm y \, \bm z} 
- \frac{1}{2 \, N_c \, C_F} 
- \frac{N_c}{2 \, C_F} \, {\hat D}_{\bm x \, \bm z} \, {\hat D}_{\bm z \, \bm w}  
+ \frac{1}{2 \, N_c \, C_F} \, {\hat D}_{\bm x \, \bm w} 
- \frac{N_c}{2 \, C_F} \, {\hat D}_{\bm u \, \bm y} \, {\hat D}_{\bm y \, \bm x}  
+ \frac{1}{2 \, N_c \, C_F} \, {\hat D}_{\bm u \, \bm x} 
\right\rangle .
\end{equation}
Note that now
\begin{subequations}
 \label{uvG}
 \begin{eqnarray}
  \bm u = \bm x + (1-\alpha) \, (\bm z - \bm x) \\ 
  \bm w = \bm x + (1-\alpha) \, (\bm y - \bm x)
 \end{eqnarray}
\end{subequations}
due to the interchanges of \eq{interchange} carried out in
Eqs.~\peq{eq-kinematic constraints}.

Separating the interaction into the symmetric and anti-symmetric
components under the ${\bm z} \leftrightarrow {\bm y}$ interchange one
obtains
\begin{subequations}\label{IGsymmanti}
\begin{eqnarray}
\mathcal{I}^{(G)}_{symm} &=& \left\langle {\hat S}_{\bm u \, \bm w} + 
\frac{N_c}{2 \, C_F} \, \left( {\hat S}_{\bm z \, \bm y} \, {\hat S}_{\bm y \, \bm z} 
- {\hat O}_{\bm z \, \bm y} \, {\hat O}_{\bm y \, \bm z}\right)
- \frac{1}{2 \, N_c \, C_F} 
- \frac{N_c}{2 \, C_F} \, \left( {\hat S}_{\bm x \, \bm z} \, {\hat S}_{\bm z \, \bm w} - {\hat O}_{\bm x \, \bm z} \, {\hat O}_{\bm z \, \bm w} \right)
+ \frac{1}{2 \, N_c \, C_F} \, {\hat S}_{\bm x \, \bm w} \right. \notag \\ && \left.
 - \frac{N_c}{2 \, C_F} \, \left( {\hat S}_{\bm u \, \bm y} \, {\hat S}_{\bm y \, \bm x} - {\hat O}_{\bm u \, \bm y} \, {\hat O}_{\bm y \, \bm x} \right) 
+ \frac{1}{2 \, N_c \, C_F} \, {\hat S}_{\bm u \, \bm x} 
\right\rangle , \\
\mathcal{I}^{(G)}_{anti} &=& i \, \left\langle {\hat O}_{\bm u \, \bm w} 
- \frac{N_c}{2 \, C_F} \, \left( {\hat S}_{\bm x \, \bm z} \, {\hat O}_{\bm z \, \bm w} + {\hat O}_{\bm x \, \bm z} \, {\hat S}_{\bm z \, \bm w} \right) 
+ \frac{1}{2 \, N_c \, C_F} \, {\hat O}_{\bm x \, \bm w} \right. \notag \\ && \left.
- \frac{N_c}{2 \, C_F} \, \left( {\hat S}_{\bm u \, \bm y} \, {\hat O}_{\bm y \, \bm x} + {\hat O}_{\bm u \, \bm y} \, {\hat S}_{\bm y \, \bm x} \right)
+ \frac{1}{2 \, N_c \, C_F} \, {\hat O}_{\bm u \, \bm x} 
\right\rangle ,
\end{eqnarray}
\end{subequations}
where we have used the fact that ${\hat O}_{\bm y \, \bm z} = - {\hat
  O}_{\bm z \, \bm y}$ which follows from the definition in \eq{Odef}.

Finally, in the large-$N_c$ limit Eqs.~\peq{IGsymmanti} simplify to
\begin{subequations}
 \label{eq-gluon production interaction}
 \begin{eqnarray}
  \mathcal{I}^{(G)}_{symm}\bigg|_{\mbox{large}-N_c} &=& S_{\bm u \, \bm w} +
  \left(S_{\bm z \, \bm y}\right)^2 - S_{\bm x \, \bm z} \, S_{\bm z \, \bm w} 
- S_{\bm u \, \bm y} \, S_{\bm y \, \bm x}  + 
  \left( O_{\bm z \, \bm y}\right)^2  + O_{\bm x \, \bm z} \, O_{\bm z \, \bm w} 
  + O_{\bm u \, \bm y} \, O_{\bm y \, \bm x}  \\
  \mathcal{I}^{(G)}_{anti}\bigg|_{\mbox{large}-N_c} &=& i \, 
  \left[ O_{\bm u \, \bm w} - S_{\bm x \, \bm z} \, 
  O_{\bm z \, \bm w} - O_{\bm x \, \bm z} \, S_{\bm z \, \bm w} -
  S_{\bm u \, \bm y} \, O_{\bm y \, \bm x} - O_{\bm u \,
  \bm y} \, S_{\bm y \, \bm x}  \right] \; .
 \end{eqnarray}
\end{subequations}
Eqs.~\peq{IGsymmanti} and \peq{Gprod}, when used in \eq{eq-Defn STSA},
give an expression for the gluon STSA in the CGC formalism. This is
another main result of this work.

Constructing the cross sections for prompt photon production out of
the gluon production cross sections we have just derived is
straightforward. One has to drop all color factors in the light-cone
wave functions, replace $\as \to \alpha_{EM} \, Z_f^2$ with $Z_f$ the
electric charge of a quark with flavor $f$ in units of the electron
charge, and recalculate the interaction with the target remembering
that the photon, in this lowest order in $\alpha_{EM}$ approximation
does not interact. One obtains the polarization-dependent and
unpolarized cross sections for photon production
\begin{subequations}
 \label{phprod}
 \begin{eqnarray}
  d(\Delta\sigma^{(\gamma)}) &=& \frac{1}{(2\pi)^3} \,  
  \int d^2 x \, d^2 y \, d^2 z
  \, e^{-i \bm k \cdot (\bm z - \bm y)} \, \Phi_{pol} (\bm x - \bm z \, , \, \bm x - \bm y, \alpha) \
  \mathcal{I}^{(\gamma)}_{anti} (\bm x \, , \, \bm y \, , \, \bm z) \label{dsigmagamma} \\
  d\sigma^{(\gamma)}_{unp} &=& \frac{1}{2 \, (2\pi)^3} \, 
  \int d^2 x \, d^2 y \, d^2 z
  \, e^{-i \bm k \cdot (\bm z - \bm y)} \, \Phi_{unp} (\bm x - \bm z \, , \, \bm x - \bm y, \alpha) \
  \mathcal{I}^{(\gamma)}_{symm} (\bm x \, , \, \bm y \, , \, \bm z) \; ,
 \end{eqnarray}
\end{subequations}
where $\Phi_{pol}$ and $\Phi_{unp}$ are given by
Eqs.~\peq{eq-polarized wavefn} and \peq{eq-unpolarized wavefn} with
the $\as \to \alpha_{EM} \, Z_f^2$ replacement.

The interaction with the target is calculated to be
\begin{equation}
  \label{Igamma}
  \mathcal{I}^{(\gamma)} = 1 + D_{\bm u \, \bm w} - D_{\bm x \, \bm w} -
  D_{\bm u \, \bm x}
\end{equation}
with the symmetric and anti-symmetric under ${\bm z} \leftrightarrow
{\bm y}$ parts
\begin{subequations}
 \label{eq-photon production interaction}
 \begin{eqnarray}
  \mathcal{I}^{(\gamma)}_{symm} &=& 1 + S_{\bm u \, \bm w}  - S_{\bm x \, \bm w} - S_{\bm u \,
  \bm x} \\
  \mathcal{I}^{(\gamma)}_{anti} &=& i \left[ O_{\bm u \, \bm w} - O_{\bm x \, \bm w} 
 - O_{\bm u \, \bm x}  \right] \; . \label{phanti}
 \end{eqnarray}
\end{subequations}
Eqs.~\peq{eq-photon production interaction} and \peq{phprod} along
with \eq{eq-Defn STSA} give us the prompt photon STSA. This is the
third and final main formal result of this work. Note that below we
will show that \eq{dsigmagamma} leads to $d(\Delta\sigma^{(\gamma)})
=0$ for any target, which implies zero STSA for photons in our
mechanism.

We have constructed general expressions for STSA generated by quark,
gluon, and photon production in $q^\uparrow + A$ collisions. Knowing
the light-cone wave functions squared \eqref{eq-unpolarized wavefn},
\eqref{eq-polarized wavefn} and the interactions for the 3 channels
\eqref{eq-symmetrized interaction} (along with \peq{direct4}),
\eqref{IGsymmanti}, \eqref{eq-photon production interaction} one can
make explicit predictions for the corresponding asymmetries. In
general terms, we have shown that in this formalism the asymmetry is
generated by coupling of the spin-dependent part of the wave function
to the Odderon interaction with the target.


\section{Evaluations and Estimates of the Asymmetry}

\label{sec-Estimates}

Unfortunately, Eqs.~\peq{eq-observables}, \peq{Gprod}, and
\peq{phprod} are too complicated to be integrated out analytically in
the general case. In this Section, in order to understand the
qualitative behavior of our results, we evaluate the integrals
analytically, taking the interaction with the target in the
quasi-classical Glauber--Mueller approximation. In such a
quasi-classical limit, the real part of the $S$ matrix \eqref{Sdef} is
\cite{Mueller:1989st}
\begin{equation}
 \label{eq-GM Pomeron}
 S_{\bm x \, \bm y} = \exp \left[ -\frac{1}{4} \, |\bm x - \bm y|^2 \ Q_s^2 \! 
\left( \frac{\bm x + \bm y}{2} \right) \, \ln \frac{1}{|\bm x - \bm y| \, \Lambda} \right],
\end{equation}
where the quark saturation scale scale $Q^2 (\bm b)$ is defined in
terms of the nuclear profile function (transverse nuclear density)
$T(\bm b)$ as
\begin{equation}
 \label{eq-saturation scale}
 Q_s^2 (\bm b) \equiv \frac{4 \, \pi \, \alpha_s^2 \, C_F}{N_c} \, T(\bm b) \; 
\end{equation} 
and $\Lambda$ is a non-perturbative IR cutoff.

In the same quasi-classical approximation the odderon amplitude is
\cite{Kovchegov:2003dm}
\begin{equation}
  \label{Ocl}
  O_{\bm x \, \bm y} = \left\langle c_0 \, \alpha_s^3 \, \ln^3 
 \frac{|\bm x - \bm r|}{|\bm y - \bm r|} \right\rangle \, 
 \exp \left[ -\frac{1}{4} \, |\bm x - \bm y|^2 \ Q_s^2 \!
\left( \frac{\bm x + \bm y}{2} \right) \, \ln \frac{1}{|\bm x - \bm y| \, \Lambda} \right]
\end{equation}
with the constant
\cite{Hatta:2005as,Kovner:2005qj,Jeon:2005cf}\footnote{Note that the
  sign is different from that in \cite{Hatta:2005as,Jeon:2005cf}: the
  sign in \eq{c0} arises when using a consistent convention for the
  sign of the coupling $g$ both in the Wilson lines and in the
  classical gluon field of the target. (Our sign convention is to have
  $+i \, g$ for the quark-gluon vertex, resulting in $+i \, g$ in the
  Wilson lines \eqref{eq-Wilson lines}.) While the physical
  conclusions reached in
  \cite{Kovchegov:2003dm,Hatta:2005as,Kovner:2005qj,Jeon:2005cf} are
  independent of the sign of the odderon amplitude, the direction of
  the asymmetry in question explicitly depends on the sign of $O_{\bm
    x \, \bm y}$.}
\begin{equation}
  \label{c0}
  c_0 = - \frac{(N_c^2 -4) \, (N_c^2 -1)}{12 \, N_c^3}.
\end{equation}
The logarithm cubed in \eq{Ocl} arises due to the triple gluon
exchange between the dipole and some quark in the target nucleus
located at transverse position $\bm r$. Angle brackets in \eq{Ocl}
denote the averaging over positions of the quark in the nuclear wave
function, along with the summation over all the nucleons in the
nucleus that may contain this quark. This averaging is carried out in
Appendix~\ref{A}, yielding
\begin{equation}
  \label{O_ave}
  O_{\bm x \, \bm y} \approx - c_0 \, \alpha_s^3 \, 
 \frac{3 \, \pi}{16} \, |\bm x - \bm y|^2 \, 
 \exp \left[ -\frac{1}{4} \, |\bm x - \bm y|^2 \ Q_s^2 \!
\left( \frac{\bm x + \bm y}{2} \right) \, \ln \frac{1}{|\bm x - \bm y| \, \Lambda} 
\right] \ (\bm x - \bm y) 
\cdot {\bm \nabla} T \! \left(\frac{\bm x + \bm y}{2} \right).
\end{equation}

For simplicity we will also work in the large-$N_c$ limit for the
light-cone wave function. Just like before, we mainly concentrate on
the quark production case in \eq{eq-observables}: STSA in the gluon
production channel can be evaluated along similar lines. We will also
consider STSA for the prompt photon production.



\subsection{Single Transverse Spin Asymmetry in Quark Production}
\label{quarkSTSAsec}

\subsubsection{Spin-Dependent Quark Production Cross Section}

First let us evaluate the numerator of the STSA in \eq{eq-Defn STSA},
which, in the quark production case, is given by \eq{dsigmaq}. (For
simplicity we assume that $\alpha < 1$ which allows us to drop the
contribution from \eq{direct3}.) Working in the large-$N_c$ limit for
the light-cone wave function we substitute the interaction from
\eq{qanti} into \eq{dsigmaq} to obtain
\begin{eqnarray}\label{dsigmaq1} 
  d(\Delta\sigma^{(q)}) = i \, \frac{N_c}{2 \, (2\pi)^3} \, \frac{\alpha}{1-\alpha} \, 
  \int d^2 x \, d^2 y \, d^2 z
  \, e^{-i \bm k \cdot (\bm z - \bm y)} \, \Phi_{pol} (\bm z - \bm x \, , \, \bm y - \bm x, \alpha) \
  && \left[ O_{\bm z \, \bm y} +
  O_{\bm
  u \, \bm w} - O_{\bm z \, \bm x} \, S_{\bm x \, \bm w} -
  O_{\bm u \, \bm x} \, S_{\bm x \, \bm y} \right. \notag \\ && \left. - S_{\bm z \,
  \bm x} \, O_{\bm x \, \bm w} - S_{\bm u \, \bm x} \,
  O_{\bm x \, \bm y} \right] . 
\end{eqnarray}
Our goal now is to evaluate this expression using the $S$-matrix from
\eq{eq-GM Pomeron} and the odderon amplitude \peq{O_ave}. 

The interaction with the target in \eq{dsigmaq1} is non-linear. It is
tempting to try to simplify the problem by neglecting all the multiple
rescattering saturation effects. In such a linearized approximation
\eq{dsigmaq1} reduces to
\begin{eqnarray}
  d(\Delta\sigma^{(q)})_{lin} = i \, \frac{N_c}{2 \, (2\pi)^3} \, 
\frac{\alpha}{1-\alpha} \, 
  \int d^2 x \, d^2 y \, d^2 z
  \, e^{-i \bm k \cdot (\bm z - \bm y)} \, \Phi_{pol} (\bm z - \bm x \, , \, \bm y - \bm x, \alpha) 
   \notag \\ \times \, \left[ o_{\bm z \, \bm y} +
  o_{\bm
  u \, \bm w} - o_{\bm z \, \bm x} -
  o_{\bm u \, \bm x} 
- o_{\bm x \, \bm w} - 
  o_{\bm x \, \bm y} \right] \label{dsigmaq_lin} 
\end{eqnarray}
where 
\begin{equation}
  \label{olin}
  o_{\bm x \, \bm y} \approx \alpha_s^3 \, 
 \frac{\pi \, N_c}{64} \, |\bm x - \bm y|^2 \, \ (\bm x - \bm y) 
\cdot {\bm \nabla} T \! \left(\frac{\bm x + \bm y}{2} \right)
\end{equation}
is the linear part of the averaged odderon amplitude \peq{O_ave}.
However, one can easily show that the cross section in
\eq{dsigmaq_lin} is in fact zero, i.e., that
\begin{equation}
  \label{linzero}
  d(\Delta\sigma^{(q)})_{lin} = 0. 
\end{equation}
We illustrate this by considering the $o_{\bm z \, \bm y}$ term in
\eq{dsigmaq_lin}. Defining new transverse vectors
\begin{equation}
  \label{tildes}
  {\tilde {\bm z}} = {\bm z} - {\bm x}, \ \ \ {\tilde {\bm y}} = {\bm y} - {\bm x},
\end{equation}
we rewrite the $o_{\bm z \, \bm y}$ contribution to the cross section
in \eq{dsigmaq_lin} as
\begin{equation}
  \label{zero1}
  i \, \frac{N_c}{2 \, (2\pi)^3} \, \frac{\alpha}{1-\alpha} \, 
  \int d^2 {\tilde y} \, d^2 {\tilde z}
  \, e^{-i \bm k \cdot ({\tilde {\bm z}} - {\tilde {\bm y}})} \, 
\Phi_{pol} ({\tilde {\bm z}}, {\tilde {\bm y}}, \alpha) \, \int d^2 x \, 
   \,  o_{{\tilde {\bm z}} + {\bm x}, \, {\tilde {\bm y}} + {\bm x}}. 
\end{equation}
This expression is zero since 
\begin{equation}
  \label{zero2}
  \int d^2 x \, 
   \,  o_{{\tilde {\bm z}} + {\bm x}, \, {\tilde {\bm y}} + {\bm x}} =0
\end{equation}
due to the fact that the odderon amplitude \peq{Odef} (and, therefore,
the linearized odderon amplitude \peq{olin}) is an anti-symmetric
function of its transverse coordinate arguments,
\begin{equation}
  \label{Oanti}
  O_{\bm x \, \bm y} = - O_{\bm y \, \bm x}.
\end{equation}
The argument goes as follows. Employing \eq{Oanti} and shifting the
integration variables we write
\begin{equation}
  \label{zero_arg}
  f ({\bm y}) \equiv \int d^2 x \ O_{{\bm x},  \, {\bm x} + {\bm y}} 
= - \int d^2 x \ O_{{\bm x},  \, {\bm x} - {\bm y}} = - f (-{\bm y}). 
\end{equation}
Since $f ({\bm y})$ depends only on one vector $\bm y$ and is a scalar
under the rotations in the transverse plane, it is a function of ${\bm
  y}^2$ only, and can satisfy \peq{zero_arg} (i.e., can be an odd
function of $\bm y$) only if $f ({\bm y}) =0$. This demonstrates that
\begin{equation}
  \label{zero_arg2}
  \int d^2 x \, O_{{\bm x},  \, {\bm x} + {\bm y}} = 0.
\end{equation}

Similar arguments can be carried out for other terms in
\eq{dsigmaq_lin}, leading in the end to \eq{linzero}. We arrive at an
important conclusion: STSA cannot result from the interaction with the
target mediated by the odderon exchange alone. Neglecting the
interactions contained in the dipole $S$-matrices in \eq{dsigmaq1}
would lead to zero transverse spin asymmetry. This is an important
observation elucidating the nature of our result \peq{eq-observables}
and the corresponding STSA: in order to generate a non-zero STSA the
interaction with the target has to contain both the $C$-odd and
$C$-even contributions!

Returning to the general case of \eq{dsigmaq1} we see that the
argument we have just presented demonstrates that the $O_{\bm z \, \bm
  y}$ and $O_{\bm u \, \bm w}$ terms are zero in the general case as
well, since they are not multiplied by the $S$-matrices. Dropping
these terms yields
\begin{eqnarray}
  d(\Delta\sigma^{(q)}) = - i \, \frac{N_c}{2 \, (2\pi)^3} \, \frac{\alpha}{1-\alpha} \, 
  \int d^2 x \, d^2 y \, d^2 z
  \, e^{-i \bm k \cdot (\bm z - \bm y)} \, \Phi_{pol} (\bm z - \bm x \, , \, \bm y - \bm x, \alpha) \
  \notag \\ \times \,  \left[ O_{\bm z \, \bm x} \, S_{\bm x \, \bm w} +
  O_{\bm u \, \bm x} \, S_{\bm x \, \bm y}  
+ O_{\bm x \, \bm w} \, S_{\bm z \, \bm x} + O_{\bm x \, \bm y} \, S_{\bm u \, \bm x} 
   \right] . \label{dsigmaq2} 
\end{eqnarray}

To evaluate \eq{dsigmaq2} let us first study its large-$k_T$
asymptotics. Since ${\tilde m} \le m$ and the quark mass $m$ is at
most the constituent quark mass of about $300$~MeV (we assume light
quark flavors), we have $k_T \gg Q_s \gg {\tilde m}$. Changing the
coordinates using \eq{tildes} reduces it to
\begin{eqnarray}
  d(\Delta\sigma^{(q)}) &=& - i \, \frac{N_c}{2 \, (2\pi)^3} \, \frac{\alpha}{1-\alpha} \, 
  \int d^2 x \, d^2 {\tilde y} \, d^2 {\tilde z}
  \, e^{-i \bm k \cdot ({\tilde {\bm z}} - {\tilde {\bm y}})} \, 
\Phi_{pol} ({\tilde {\bm z}} \, , \, {\tilde {\bm y}}, \alpha) \
  \notag \\ & \times & \,  \left[ O_{\bm x + {\tilde {\bm z}},  \, \bm x} \ S_{\bm x, \, \bm x + \alpha \, {\tilde {\bm y}}} +
  O_{\bm x + \alpha \, {\tilde {\bm z}}, \ \bm x} \ S_{\bm x, \, \bm x + {\tilde {\bm y}} }  
+ O_{\bm x, \,  \bm x + \alpha \, {\tilde {\bm y}}} \ S_{\bm x, \, \bm x + {\tilde {\bm z}}} + O_{\bm x, \, \bm x + {\tilde {\bm y}} } \ S_{\bm x , \, \bm x + \alpha \, {\tilde {\bm z}}} 
   \right] . \label{dsigmaq3} 
\end{eqnarray}
For each term in the square brackets of \eq{dsigmaq3} the integrals
over ${\tilde {\bm z}}$ and ${\tilde {\bm y}}$ factorize: taking the
large-$k_T$ limit in each of them separately, we see that the
large-$k_T$ asymptotics corresponds to small ${\tilde {z}}_T$ and
${\tilde {y}}_T$. We thus need to expand the interaction with the
target term in the square brackets of \eq{dsigmaq3} to the lowest
non-trivial order in ${\tilde {z}}_T$ and ${\tilde {y}}_T$. Note that
above we have seen that if we keep the dipole $S$-matrices at the
lowest order in the dipole size, $S=1$, then the spin-dependent cross
section would be zero. We thus use Eqs.~\peq{eq-GM Pomeron} and
\peq{O_ave} to expand the $S$-matrices to the next-to-lowest order,
while keeping the odderon amplitudes at the lowest order given by
\eq{olin}.  Performing the expansion, substituting the wave function
squared from \eq{eq-polarized wavefn} (also expanded to the lowest
non-trivial order in ${\tilde {z}}_T$ and ${\tilde {y}}_T$) into
\eq{dsigmaq3}, and employing \eq{eq-saturation scale} we obtain
\begin{eqnarray}
  d && (\Delta\sigma^{(q)}) \bigg|_{k_T \gg Q_s} \approx i \, 
\frac{N_c^2}{1024 \, \pi^2} \, \alpha_s^6 \,  {\tilde m} \, \alpha^4 \,
  \int d^2 x \, d^2 {\tilde y} \, d^2 {\tilde z}
  \, e^{-i \bm k \cdot ({\tilde {\bm z}} - {\tilde {\bm y}})} \, 
  \bigg(
 \frac{{\tilde z}^{2} }{{\tilde z}_T^2} \, \ln \frac{1}{\tilde m \, {\tilde y}_T} 
  + \, \frac{{\tilde y}^{2}}{{\tilde y}_T^2} \, \ln \frac{1}{\tilde m \, {\tilde z}_T} \bigg) 
\, {\tilde z}_T^2 \, {\tilde y}_T^2
\notag \\ && \times \, \left[ {\tilde {\bm z}} \cdot {\bm \nabla} 
T \! \left({\bm x} + \frac{{\tilde {\bm z}}}{2} \right) \ 
T \! \left({\bm x} + \frac{\alpha \, {\tilde {\bm y}}}{2} \right) \, 
\ln \frac{1}{\alpha \, {\tilde y}_T \, \Lambda} + 
\alpha \ {\tilde {\bm z}} \cdot {\bm \nabla} 
T \! \left({\bm x} + \frac{\alpha \, {\tilde {\bm z}}}{2} \right) \ 
T \! \left({\bm x} + \frac{{\tilde {\bm y}}}{2} \right) \, 
\ln \frac{1}{{\tilde y}_T \, \Lambda} 
- ({\tilde {\bm z}} \leftrightarrow {\tilde {\bm y}}) \right] 
. \label{dsigmaq4} 
\end{eqnarray}

Since ${\tilde {z}}_T$ and ${\tilde {y}}_T$ are small, one may think
of neglecting them compared to $\bm x$ in the arguments of $T$'s in
\eq{dsigmaq4}. However, this would lead to a zero answer after
integration over $\bm x$.  The reason for this conclusion is that any
unpolarized target, after averaging over many events, is rotationally
symmetric in the transverse plane. This implies that ${\bm \nabla} T
({\bm x}) = {\bm \nabla} T (x_T) = {\hat x} \, T' (x_T)$ where ${\hat
  x}$ is a unit vector in the direction of $\bm x$ and $T' (x_T) = d
T(x_T)/d x_T$. Integrating ${\hat x}$ over the angles of $\bm x$ would
give zero.

Instead of neglecting ${\tilde {z}}_T$ and ${\tilde {y}}_T$, we shift
${\bm x} \to {\bm x} - {\tilde {\bm z}}/2$ in the first term in the
square brackets of \eq{dsigmaq4} and expand $T$ along the lines of
\eq{expansion}, and perform similar operations to the other terms in
the brackets obtaining
\begin{eqnarray}
  d && (\Delta\sigma^{(q)}) \bigg|_{k_T \gg Q_s} \approx i \, 
\frac{N_c^2}{2048 \, \pi^2} \, \alpha_s^6 \,  {\tilde m} \, \alpha^4 \,
  \int d^2 x \, d^2 {y} \, d^2 {z}
  \, e^{-i \bm k \cdot ({{\bm z}} - {{\bm y}})} \, 
  \bigg(
 \frac{{z}^{2} }{{z}_T^2} \, \ln \frac{1}{\tilde m \, {y}_T} 
  + \, \frac{{y}^{2}}{{y}_T^2} \, \ln \frac{1}{\tilde m \, {z}_T} \bigg) 
\, {z}_T^2 \, {y}_T^2
\notag \\ && \times \, \left[ {{\bm z}} \cdot {\bm \nabla} 
T \! \left({\bm x} \right) \ (\alpha \, {\bm y} - {\bm z}) \cdot {\bm \nabla} 
T \! \left({\bm x} \right) \, 
\ln \frac{1}{\alpha \, y_T \, \Lambda} + 
\alpha \ {{\bm z}} \cdot {\bm \nabla} 
T \! \left({\bm x}\right) \ ({\bm y} - \alpha \, {\bm z}) \cdot {\bm \nabla} 
T \! \left({\bm x}\right) \, 
\ln \frac{1}{y_T \, \Lambda} - ({{\bm z}} \leftrightarrow {{\bm y}}) \right] 
, \label{dsigmaq5} 
\end{eqnarray}
where we have dropped the tildes over $\bm y$ and $\bm z$, since now
it would not cause confusion.

Using ${\bm \nabla} T ({\bm x}) = {\bm \nabla} T (x_T) = {\hat x} \,
T' (x_T)$ and integrating over the angles of $\bm x$ reduces
\eq{dsigmaq5} to
\begin{eqnarray}
  d && (\Delta\sigma^{(q)}) \bigg|_{k_T \gg Q_s} \approx i \, 
\frac{N_c^2}{4096 \, \pi} \, \alpha_s^6 \,  {\tilde m} \, \alpha^4 \,
  \int\limits_0^\infty  d x_T^2 \, [T' (x_T)]^2 \, \int d^2 {y} \, d^2 {z}
  \, e^{-i \bm k \cdot ({{\bm z}} - {{\bm y}})} \, 
  \bigg(
 \frac{{z}^{2} }{{z}_T^2} \, \ln \frac{1}{\tilde m \, {y}_T} 
  + \, \frac{{y}^{2}}{{y}_T^2} \, \ln \frac{1}{\tilde m \, {z}_T} \bigg) 
\, {z}_T^2 \, {y}_T^2
\notag \\ && \times \, \left[ {{\bm z}} \cdot  (\alpha \, {\bm y} - {\bm z})  \, 
\ln \frac{1}{\alpha \, y_T \, \Lambda} + 
\alpha \ {{\bm z}} \cdot  ({\bm y} - \alpha \, {\bm z}) \, 
\ln \frac{1}{y_T \, \Lambda} 
- 
\alpha \ {{\bm y}} \cdot  ({\bm z} - \alpha \, {\bm y})  \, 
\ln \frac{1}{z_T \, \Lambda} - {{\bm y}} \cdot (\alpha \, {\bm z} -  {\bm y}) \, 
\ln \frac{1}{\alpha \, z_T \, \Lambda} \right]. \label{dsigmaq6} 
\end{eqnarray}
Integrating over $\bm y$ and $\bm z$ in \eq{dsigmaq6} and discarding
delta-functions of $\bm k$ (since $k_T \neq 0$) yields
\begin{eqnarray}
  d (\Delta\sigma^{(q)}) \bigg|_{k_T \gg Q_s} \approx && \  
\frac{\pi \, N_c^2}{8} \, \alpha_s^6 \,  {\tilde m} \, 
\alpha^4 \, (2 + 3 \, \alpha + 2 \, \alpha^2) \, 
  \int\limits_0^\infty  d x_T^2 \, [T' (x_T)]^2 \, \frac{k^2}{k_T^{10}}. \label{dsigmaq7} 
\end{eqnarray}
We see that the polarized spectrum falls off rather steeply with
$k_T$, scaling as $1/k_T^9$. This indicates that in the standard
collinear factorization framework our STSA generating mechanism
originates in some higher-twist operator. 

Another important qualitative feature one can see in \eq{dsigmaq7} is
that the spin-dependent cross section falls off with decreasing
longitudinal momentum fraction $\alpha$, which implies that the
corresponding STSA decreases with decreasing Feynman-$x$ of the
projectile, in qualitative agreement with the experimental data.

To improve on \eq{dsigmaq7} let us find the spin-dependent
differential cross section $d (\Delta\sigma^{(q)})$ for lower $k_T$,
closer to the saturation scale. To be more specific let us relax the
$k_T \gg Q_s$ restriction and consider a broader region of $k_T
\lesssim Q_s$ and $k_T \gtrsim Q_s$, but still with $k_T \gg {\tilde
  m}$.  For such not very large $k_T$ we can neglect the logarithms in
the exponents of Eqs.~\peq{eq-GM Pomeron} and \peq{O_ave} as slowly
varying functions compared to powers they multiply
\cite{Kovchegov:1998bi,Kharzeev:2003wz,Jalilian-Marian:2005jf},
writing
\begin{equation}
 \label{Sapp}
 S_{\bm x \, \bm y} \approx \exp \left[ -\frac{1}{4} \, |\bm x - \bm y|^2 \ Q_s^2 \! 
\left( \frac{\bm x + \bm y}{2} \right) \right]
\end{equation}
and
\begin{equation}
  \label{Oapp}
  O_{\bm x \, \bm y} \approx - c_0 \, \alpha_s^3 \, 
 \frac{3 \, \pi}{16} \, |\bm x - \bm y|^2 \, 
 \exp \left[ -\frac{1}{4} \, |\bm x - \bm y|^2 \ Q_s^2 \!
\left( \frac{\bm x + \bm y}{2} \right) \right] \ (\bm x - \bm y) 
\cdot {\bm \nabla} T \! \left(\frac{\bm x + \bm y}{2} \right).
\end{equation}
Substituting Eqs.~\peq{Sapp} and \peq{Oapp} into \eq{dsigmaq3},
expanding the polarized wave function squared, and dropping the tildes
yields
\begin{eqnarray}
  && d(\Delta\sigma^{(q)}) \approx 
- i \, \frac{N_c^2}{512 \, \pi^3} \, 
\as^4 \, {\tilde m} \, \alpha^2 \, 
  \int d^2 x \, d^2 y \, d^2 z
  \, e^{-i \bm k \cdot ({{\bm z}} - {{\bm y}})} \, 
\bigg(
 \frac{{z}^{2} }{{z}_T^2} \, \ln \frac{1}{\tilde m \, {y}_T} 
  + \, \frac{{y}^{2}}{{y}_T^2} \, \ln \frac{1}{\tilde m \, {z}_T} \bigg)
   \,  \bigg[ z_T^2 \, {\bm z} \cdot {\bm \nabla} 
T \! \left({\bm x} + \frac{{\bm z}}{2} \right) \notag \\ & \times & \, 
e^{-\frac{1}{4} \, z_T^2 \, Q_s^2 \left({\bm x} + \frac{{\bm z}}{2} \right)
-\frac{1}{4} \, \alpha^2 \, y_T^2 \, 
Q_s^2 \left({\bm x} + \frac{\alpha \, {\bm y}}{2} \right)}
+ \alpha^3 \, z_T^2 \, {\bm z} \cdot {\bm \nabla} 
T \! \left({\bm x} + \frac{\alpha \, {\bm z}}{2} \right) \, 
e^{-\frac{1}{4} \, \alpha^2 \, z_T^2 \, 
Q_s^2 \left({\bm x} + \frac{\alpha \, {\bm z}}{2} \right)
-\frac{1}{4} \, y_T^2 \, Q_s^2 \left({\bm x} + \frac{{\bm y}}{2} \right)}
- ({\bm z} \leftrightarrow {\bm y}) \bigg]
. \label{dsigmaq8} 
\end{eqnarray}
Similar to the large-$k_T$ asymptotics, we shift ${\bm x} \to {\bm x}
- {\bm z}/2$ in the first term in the square brackets of \eq{dsigmaq8}
and expand the resulting exponential with the help of
\eq{eq-saturation scale} as
\begin{equation}
  \label{Expansion}
  e^{-\frac{1}{4} \, z_T^2 \, Q_s^2 ({\bm x})
-\frac{1}{4} \, \alpha^2 \, y_T^2 \, 
Q_s^2 \left({\bm x} + \frac{\alpha \, {\bm y} - {\bm z}}{2} \right)} 
\approx \left[1 - 
\frac{\pi}{4} \, \as^2 \, \alpha^2 \, y_T^2 \, 
(\alpha \, {\bm y} - {\bm z}) \cdot {\bm \nabla} T ({\bm x}) \right] \, 
e^{-\frac{1}{4} \, z_T^2 \, Q_s^2 ({\bm x})
-\frac{1}{4} \, \alpha^2 \, y_T^2 \, Q_s^2 ({\bm x})}.
\end{equation}
The $1$ in the square brackets of \eq{Expansion} does not contribute
as its contribution vanishes after integration over the angles of $\bm
x$ in \eq{dsigmaq8}, leaving only the second term to contribute.
Performing similar expansions in the other terms in the square
brackets of \eq{dsigmaq8} we obtain
\begin{eqnarray}
  && d(\Delta\sigma^{(q)}) \approx 
i \, \frac{N_c^2}{2048 \, \pi^2} \, 
\as^6 \, {\tilde m} \, \alpha^4 \, 
  \int d^2 x \, d^2 y \, d^2 z
  \, e^{-i \bm k \cdot ({{\bm z}} - {{\bm y}})} \, 
\bigg(
 \frac{{z}^{2} }{{z}_T^2} 
  + \, \frac{{y}^{2}}{{y}_T^2} \bigg) \, z_T^2 \, y_T^2 
    \, \bigg[ {\bm z} \cdot {\bm \nabla} 
T ({\bm x})  \
(\alpha \, {\bm y} - {\bm z}) \cdot {\bm \nabla} T ({\bm x})  \notag \\ & \times & \,
e^{-\frac{1}{4} \, z_T^2 \, Q_s^2 ({\bm x})
-\frac{1}{4} \, \alpha^2 \, y_T^2 \, 
Q_s^2 ({\bm x})}
+ \alpha \, {\bm z} \cdot {\bm \nabla} 
T ({\bm x}) \ ({\bm y} - \alpha \, {\bm z}) \cdot {\bm \nabla} 
T ({\bm x}) \,
e^{-\frac{1}{4} \, \alpha^2 \, z_T^2 \, 
Q_s^2 ({\bm x}) - \frac{1}{4} \, y_T^2 \, Q_s^2 ({\bm x})}
- ({\bm z} \leftrightarrow {\bm y}) \bigg], \label{dsigmaq9} 
\end{eqnarray}
where we have also dropped $\ln 1/{\tilde m} \, {y}_T$ and $\ln
1/{\tilde m} \, {z}_T$, since, with our precision, similar logarithms
were neglected in Eqs.~\peq{Sapp} and \peq{Oapp} above as slowly
varying functions of their arguments \footnote{We have done the
  calculation without neglecting those logarithms: the resulting
  changes were mainly of quantitative nature, while the obtained
  expression was significantly more complicated than \eq{dsigmaq11}.
  Since both the expressions with and without the logarithms are
  approximate, we decided to only show the latter in this work due to
  its relative compactness.}. Again, integrating over the angles of
$\bm x$ yields
\begin{eqnarray}
  && d(\Delta\sigma^{(q)}) \approx 
i \, \frac{N_c^2}{4096 \, \pi } \, 
\as^6 \, {\tilde m} \, \alpha^4 \, 
  \int\limits_0^\infty  d x_T^2 \, [T' (x_T)]^2 \, \int d^2 y \, d^2 z
  \, e^{-i \bm k \cdot ({{\bm z}} - {{\bm y}})} \, 
\bigg(\frac{{z}^{2} }{{z}_T^2} + \, \frac{{y}^{2}}{{y}_T^2} \bigg) \, z_T^2 \, y_T^2 
   \notag \\ & \times & \,  \bigg[  
(\alpha^2 \, y_T^2 - z_T^2)  \,
e^{-\frac{1}{4} \, z_T^2 \, Q_s^2 ({x}_T)
-\frac{1}{4} \, \alpha^2 \, y_T^2 \, 
Q_s^2 ({x}_T)}
+ (y_T^2 - \alpha^2 \, z_T^2) \,
e^{-\frac{1}{4} \, \alpha^2 \, z_T^2 \, 
Q_s^2 ({x}_T) - \frac{1}{4} \, y_T^2 \, Q_s^2 ({x}_T)}
\bigg]. \label{dsigmaq10} 
\end{eqnarray}
Integrating over $\bm y$ and $\bm z$ we get
\begin{eqnarray}
  d(\Delta\sigma^{(q)}) \approx  
\frac{\pi \, N_c^2}{4} \, 
\frac{\as^6 \, {\tilde m}}{\alpha^4}  
  \int\limits_0^\infty  d x_T^2 \, [T' (x_T)]^2 
\, \frac{k^2 \, k_T^2}{Q_s^{14} ({x}_T)} \,
\left[ (1 - \alpha^2)^2 \, k_T^2 - \alpha^2 \, (1 + \alpha^2) \, Q_s^2 ({x}_T) \right]  \,
e^{-\frac{k_T^2}{Q_s^2 ({x}_T)} \, \left( 1 + \frac{1}{\alpha^2} \right)}.  
\label{dsigmaq11} 
\end{eqnarray}

This is the final expression for the STSA-generating cross section for
quark production. Note again that $k^2$ is the $y$-component of the
quark's transverse momentum ${\bm k} = (k^1, k^2)$. Let us point out a
few of the important features of \eq{dsigmaq11}.  First of all, we see
that similar to \eq{dsigmaq7} decreases with decreasing $\alpha$ for
small $\alpha$, now due to the factor of $1/\alpha^2$ in the exponent.
We also see that for $k_T \to 0$ the cross section
$d(\Delta\sigma^{(q)})$ also goes to zero. We also note that the cross
section \peq{dsigmaq11} is not a monotonic function of $k_T$. In
particular, for positive $k^2$ it starts out negative at small $k_T$,
becoming positive for $k_T > Q_s \, \alpha \,
\sqrt{1+\alpha^2}/(1-\alpha^2)$, in agreement with the large-$k_T$
asymptotics of \eq{dsigmaq7}.


\subsubsection{The Unpolarized Cross-Section}
 
\label{subsec-dsigmaunp}

The real hadronic STSA in \eq{eq-Defn STSA} contains contributions
from both quark and gluon production cross sections
\peq{eq-observables} and \peq{Gprod} in the numerator and in the
denominator, convoluted with the fragmentation functions for the
quarks and gluons decaying into a particular hadron species as well as
the transversity distribution of polarized quarks. This is what needs
to be done to have a real comparison of the data with our theoretical
results. While such comparison is beyond the scope of this work, we
would like to assess the main qualitative features of our
STSA-generating mechanism by concentrating on quark STSA only.

It may be tempting to consider a situation where both the numerator
and the denominator of \eq{eq-Defn STSA} are driven by the quark
contributions. However, the unpolarized valence quark production cross
section \peq{dsigmaq_unp} is known to decrease with decreasing quark
momentum fraction $\alpha$ \cite{Itakura:2003jp,Albacete:2006vv},
while both the unpolarized gluon and sea quark production cross
sections grow with decreasing $\alpha$ in theoretical calculations
\cite{Kovchegov:2001sc,Kovchegov:2006qn}. In the actual experiments
the hadron multiplicity also increases as we move further away from
the projectile in rapidity.

Therefore, in order to get a somewhat realistic evaluation of the
qualitative behavior of the obtained STSA, we will use the unpolarized
gluon production cross section in the denominator of \eq{eq-Defn
  STSA}. While evaluation of the unpolarized gluon cross section
\peq{Gunp} along the same lines as were used to obtain \eq{dsigmaq11}
is somewhat involved, we will approximate the result by assuming that
the produced gluon is soft (i.e., far from the projectile in
rapidity), in which case the corresponding production cross section is
\cite{Kovchegov:1998bi,Kharzeev:2003wz,Jalilian-Marian:2005jf}
\begin{eqnarray}
 \label{Gunp2}
 d\sigma_{unp}^{(G)} \approx \frac{\alpha_s \, N_c}{2 \, \pi} \, 
 \, \int\limits_0^\infty dx_T^2 \, \bigg\{ - \frac{1}{k_T^2} 
+ \frac{2}{k_T^2} \, e^{-\frac{k_T^2}{Q_s^2 (x_T)}} +
 \frac{1}{Q_s^2 (x_T)} \, e^{-\frac{k_T^2}{Q_s^2 (x_T)}} \,
 \bigg[ \mathrm{Ei}\bigg(\frac{k_T^2}{Q_s^2 (x_T)}\bigg) - 
\ln\frac{4 \, k_T^2 \, \Lambda^2}{Q_s^4 (x_T)} \bigg] \bigg\}.
\end{eqnarray}

\subsubsection{Single Transverse Spin Asymmetry}


We now have all the essential ingredients to sketch the STSA due to
quark production in the large-$N_c$ limit (for the wave function): we
have Eqs.~\peq{dsigmaq11} and \peq{Gunp2}, giving the numerator and
the denominator of \eq{eq-Defn STSA} correspondingly. We thus write
\begin{eqnarray}
  A_N^{(q)} ({\bm k}) &=& \frac{\pi^2 \, N_c \, \as^5 \, m}{4} \, 
\frac{1-\alpha}{\alpha^4} \, 
\int\limits_0^\infty  d x_T^2 \, [T' (x_T)]^2 
\, \frac{k^2 \, k_T^2}{Q_s^{14} ({x}_T)} \,
\left[ (1 - \alpha^2)^2 \, k_T^2 - \alpha^2 \, (1 + \alpha^2) \, Q_s^2 ({x}_T) \right]  \,
e^{-\frac{k_T^2}{Q_s^2 ({x}_T)} \, \left( 1 + \frac{1}{\alpha^2} \right)} \notag \\ 
&\times& \, \Bigg[ \int\limits_0^\infty dy_T^2 \, \bigg\{ - \frac{1}{k_T^2} 
+ \frac{2}{k_T^2} \, e^{-\frac{k_T^2}{Q_s^2 (y_T)}} +
 \frac{1}{Q_s^2 (y_T)} \, e^{-\frac{k_T^2}{Q_s^2 (y_T)}} \,
 \bigg[ \mathrm{Ei}\bigg(\frac{k_T^2}{Q_s^2 (y_T)}\bigg) - 
\ln\frac{4 \, k_T^2 \, \Lambda^2}{Q_s^4 (y_T)} \bigg] \bigg\} \Bigg]^{-1} . \label{ANq}
\end{eqnarray}
The $x_T$- and $y_T$-integrals in \eq{ANq} appear to be very hard to
evaluate analytically. Instead we evaluate the integrals numerically
assuming a simple Gaussian form of the nuclear profile function,
\begin{equation}
  \label{Tgauss}
  T (\bm b) = \frac{4}{3} \, R \, \rho \ e^{-b_T^2/R^2}
\end{equation}
with $R$ the nuclear radius and $\rho$ the nucleon density. Such
Gaussian profiles are of course not realistic for nuclei, but have
been successfully used to describe protons (see e.g.
\cite{Ayala:1996em}).

In evaluating the STSA in \eq{ANq} one has to remember that in the
standard convention one has to choose $\bm k$ in the direction left of
the beam, which, in our notation, means along the negative $y$-axis
(see \fig{orientation}). Hence we need to replace $k^2 \to - k_T$ in
\eq{ANq}.

\begin{figure}[h]
  \includegraphics[width=10cm]{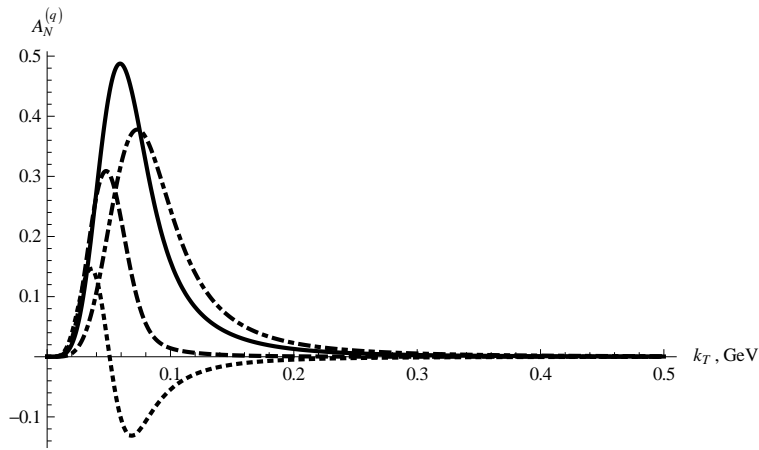}
 \caption{Quark STSA from \eq{ANq} for the proton target plotted as 
   a function of $k_T$ for different values of the longitudinal
   momentum fraction $\alpha$ carried by the produced quark: $\alpha =
   0.9$ (dash-dotted curve), $\alpha = 0.7$ (solid curve), $\alpha =
   0.6$ (dashed curve), and $\alpha = 0.5$ (dotted curve).}
\label{AN} 
\end{figure}

To plot \eq{ANq} we will attempt to use somewhat realistic numbers,
while realizing that all the theoretically-calculated cross sections
are likely to have non-perturbative normalization corrections, which
may affect the size of the effect. To that end, we will use the
saturation scale (cf. \eq{eq-saturation scale})
\begin{equation}
  \label{Qsat_real}
  Q_s^2 ({\bm b}) = 2 \, \pi \, \as^2 \, K^2 \, T ({\bm b})
\end{equation}
with the $K$-factor fixed at $K=10$ to make $Q_s \approx 1$~GeV, which
is a realistic value for a proton. (Each $T' (x_T)$ in \eq{ANq} is
multiplied by the same $K^2$-factor, since it also arises from the
saturation scale.) We put $m=300$~MeV to mimic a constituent quark,
along with $\rho = 0.35$~fm$^{-3}$ for a proton of radius $R =
0.878$~fm, and $\as =0.3$.  We plot the resulting $A_N^{(q)}$ from
\eq{ANq} in \fig{AN} for different values of $\alpha$ with the IR
cutoff $\Lambda = 100$~MeV and cutting off the $x_T$- and $y_T$
integrals in \eq{ANq} at $2.1$~fm in the IR. (Note that
strictly-speaking the CGC formalism employed here is valid only for
scattering on a nuclear target, since it resums powers of a large
parameter $\as^2 \, A^{1/3}$. However its applications to proton
target have been successful phenomenologically in the past
\cite{Albacete:2010sy}, giving one hope that our estimates here could
be relevant for $p^{\uparrow} + p$ collisions.)

From \fig{AN} we see that our STSA is a non-monotonic function of
transverse momentum $k_T$, first rising and then falling off with
$k_T$ in qualitative agreement with the data shown in the right panel
of \fig{fig-Experimental Data}. As one can clearly see from \eq{ANq}
the maximum of $A_N^{(q)}$ at impact parameter $x_T$ in our formalism
is determined (up to a constant) by the saturation scale, $k_T \sim
Q_s (x_T)$, such that the asymmetry integrated over all impact
parameters peaks at $k_T \sim Q_s$ with $Q_s$ an effective averaged
saturation scale. The conclusion about $A_N$ peaking at $k_T \approx
Q_s$ was previously reached in \cite{Boer:2006rj}. Let us stress again
that the STSA in our case changes sign when plotted as a function of
$k_T$ or $\alpha$ (i.e., it has a ``node'').


Note that, while the magnitude of STSA plotted in \fig{AN} can be as
large as tens of percent, like the data in \fig{fig-Experimental
  Data}, the momentum at which the asymmetry is non-zero appears to be
much smaller in our \fig{AN} than it is in the data of
\fig{fig-Experimental Data}. The discrepancy of the $k_T$-range of the
data and our \fig{AN} signals the following potential problem: the
$x_T$-integral in \eq{ANq} is dominated by large $x_T$, where $Q_s
(x_T)$ is small, leading to small values of $k_T$ dominating $A_N$,
and potentially making the corresponding physics
non-perturbative. Thus our perturbative calculation appears to be
sensitive to the non-perturbative domain.

\begin{figure}[h]
  \includegraphics[width=10cm]{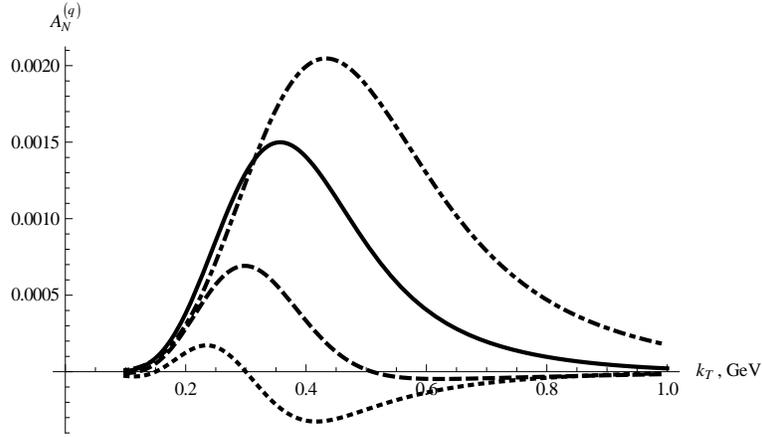}
  \caption{Same as in \fig{AN}, but with the $1.3$~fm upper cutoff on
    the $x_T$- and $y_T$ integrals in \eq{ANq} and a factor of $\ln
    k_T / {\tilde m}$ inserted.}
\label{AN2} 
\end{figure}

To illustrate the range of spectra that can be obtained by our
estimates, we replot $A_N$ from \fig{AN} in \fig{AN2} cutting off the
$x_T$- and $y_T$-integrals in \eq{ANq} by $1.3$~fm.  In addition, we
mimic the coordinate-space logarithms, like those that were neglected
after \eqref{dsigmaq8}, by introducing a factor of $\ln k_T / {\tilde
  m}$.  In this plot the $k_T$-range of the asymmetry is broader than
in \fig{AN}, which makes it closer to the experimental data in
\fig{fig-Experimental Data}, but the height of the asymmetry is over
an order-of-magnitude lower than the data.  More work is needed to
assess whether the cutoff dependence is a result of the approximations
made, or whether it actually signals a potential breakdown of the
approach indicating the non-perturbative nature of STSA.

Another important observation one can make from \fig{AN} is that
$A_N^{(q)}$ increases with increasing $\alpha$, except for very large
values of $\alpha$ when it starts to decrease. The increase of
$A_N^{(q)}$ with increasing $\alpha$ is in qualitative agreement with
the data in the left panel of \fig{fig-Experimental Data}, where the
data points increase with increasing Feynman-$x$. One can also see
that while our STSA in \fig{AN} is mostly positive, the data for
$\pi^-$ mesons in the left panel of \fig{fig-Experimental Data} gives
a negative STSA. This result can be explained in the following simple
model. Imagine a constituent-quark model of a proton, with the spins
of both up quarks aligned with the net proton spin, and the spin of
the down quark pointing in the opposite direction. Then in our
mechanism each up-quark would give a positive STSA denoted $A_u$,
while the down-quark would give a negative STSA of the same absolute
value, denoted $A_d = -A_u$. Imagining that after the collision the
up-quarks fragment into $\pi^+$'s and $\pi^0$'s, while the down quarks
fragment into $\pi^0$'s and $\pi^-$'s, and taking into account that
there are twice as many up-quarks than down-quarks in the proton, and
neglecting pions resulting from the gluon fragmentation, we obtain
$A_{\pi^+} = A_u$, $A_{\pi^0} = A_u/2$, and $A_{\pi^-} = A_d/2 = -
A_u/2$, thus obtaining a negative STSA for $\pi^-$'s.\footnote{We
  would like to thank Mickey Chiu for a discussion of these
  estimates.} This naive model appears to be in a qualitative
agreement (and in loose quantitative agreement) with the data in the
left panel of \fig{fig-Experimental Data}.  Inclusion of gluon
fragmentation may further improve the quantitative agreement with the
data.

\begin{figure}[h]
  \includegraphics[width=10cm]{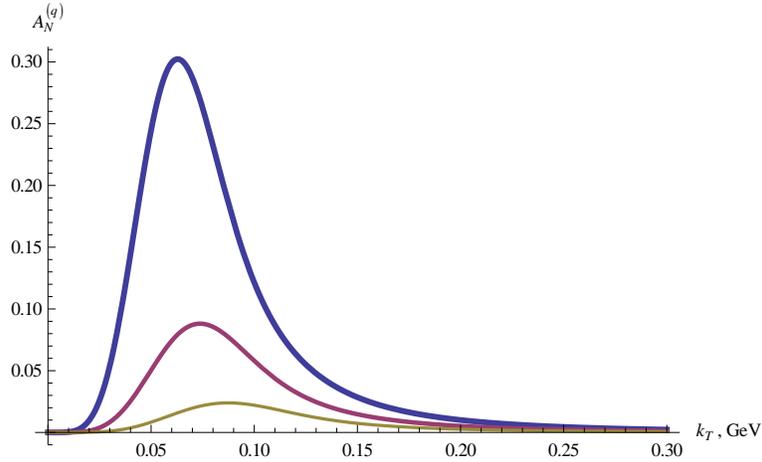}
 \caption{Quark STSA from \eq{ANq} plotted as a function of $k_T$ for different 
   values of the target radius: $R = 1$~fm (top curve), $R = 1.4$~fm
   (middle curve), and $R = 2$~fm (bottom curve) for $\alpha = 0.7$.}
\label{ANrad} 
\end{figure}

Finally, to test the dependence of our STSA in \eq{ANq} on the size of
the target, we note that for $k_T \approx Q_s$ one gets
\begin{equation}
  \label{Adep}
  A_N^{(q)} (k_T \approx Q_s) \sim \frac{1}{Q_s^7} \sim A^{-7/6},
\end{equation}
if $Q_s^2 \sim A^{1/3}$. This indicates a very steep falloff of STSA
with the atomic number of the nuclear target. Such a conclusion
appears to be supported by the numerical evaluation of \eq{ANq} for
several different radii of the target shown in \fig{ANrad}. (Now the
$x_T$- and $y_T$ integrals are cut off at $2.4$~fm.) One can see that
$A_N^{(q)}$ drops very rapidly with the size of the target.
If the experimentally observed STSA in $p^{\uparrow} + p$ collisions
are due to our mechanism, our prediction is then that in $p^{\uparrow}
+ A$ collisions STSA should be much smaller than that in $p^{\uparrow}
+ p$. In the case of a heavy ion target like $Au$ the STSA due to our
mechanism is likely to be negligibly small.

While we have demonstrated here the potential for our calculations to
agree with the data, the evaluations presented here have to be
significantly improved to reach a definitive conclusion. For instance
the $k_T$-dependence and the overall normalization in our
Eqs.~\peq{dsigmaq11} and \peq{Gunp2} are overly simple and is not
ready to be compared to the data. The equations need to be corrected
for the effects of DGLAP evolution, small-$x$ evolution and for the
running of the coupling for a meaningful quantitative comparison with
the data.  Only such a phenomenological analysis can determine whether
our mechanism for generating STSA is dominant, or whether it is simply
one of the many factors contributing to the asymmetry.



\subsection{STSA in Photon Production}

\label{subsec-Photon STSA (est)} 

Using the methods developed in Sec.~\ref{quarkSTSAsec} we can now
evaluate the photon STSA given by Eqs.~\peq{phprod} and \peq{eq-photon
  production interaction}. Substituting \eq{phanti} into
\eq{dsigmagamma} and performing the variable shift of \eq{tildes}
while keeping in mind that now $\bm u$ and $\bm w$ are given by
Eqs.~\peq{uvG} yields
\begin{equation}
  \label{dsigmaph1}
  d(\Delta\sigma^{(\gamma)}) = \frac{i}{(2\pi)^3} \,  
  \int d^2 x \, d^2 y \, d^2 z
  \, e^{-i \bm k \cdot (\bm z - \bm y)} \, \Phi_{pol} (- \bm z \, , \, - \bm y, \alpha) \
  \left[ O_{{\bm x} + (1-\alpha) \, {\bm z},  \, {\bm x} + (1-\alpha) \, {\bm y}} 
- O_{{\bm x} , \, {\bm x} + (1-\alpha) \, {\bm y}} 
- O_{{\bm x} + (1-\alpha) \, {\bm z}, \, {\bm x}} \right]
\end{equation}
where we have again dropped the tildes for brevity. Using the argument
of Eqs.~\peq{zero_arg} and \peq{zero_arg2} we see that each term in
the square brackets in \eq{dsigmaph1} is zero after the integration
over $\bm x$. We thus have an exact result that
\begin{equation}
  \label{dsigmaph2}
  d(\Delta\sigma^{(\gamma)}) = 0
\end{equation}
in our mechanism for generating photon STSA. Hence the photon STSA is
zero, $A_N^{(\gamma)} =0$, in the forward production region under
consideration.



\section{Conclusions}

\label{sec-Concl} 

To conclude let us summarize the main points of this work. Above we
have shown how STSA can be generated in the CGC formalism for quark
and gluon production. The results for the corresponding cross sections
are given in Eqs.~\peq{eq-observables}, \peq{direct3} and \peq{Gprod}.
The same mechanism gives zero STSA for prompt photons.

In our case STSA is generated by both a splitting in the projectile
wave function, and by the combination of the $C$-odd and $C$-even
interactions with the target. Hence our STSA-generating mechanism is
distinctively different from the Collins \cite{Collins:1992kk} and
Sivers \cite{Sivers:1989cc,Sivers:1990fh} effects, and is more akin to
(though still different from) the higher-twist mechanisms of
\cite{Efremov:1981sh,Efremov:1984ip,Qiu:1991pp,Ji:1992eu,Qiu:1998ia,Brodsky:2002cx,Collins:2002kn,Koike:2011mb,Kanazawa:2000hz,Kanazawa:2000kp}.

Evaluating the quark STSA in a simplified quasi-classical model we
found qualitative agreement with the data: quark STSA appears to be a
non-monotonic function of $k_T$, and is an increasing function of
increasing $x_F$ (for most of the $x_F$-range). It is perhaps
encouraging that the obtained asymmetry can be of the
order-of-magnitude of the experimental data. Further phenomenological
studies of our formulas \peq{eq-observables} and \peq{Gprod} are
needed to determine whether this qualitative agreement of our results
with the data can become quantitative.

Analyzing the general quark production formula \peq{eq-cross section}
one can see that the contribution to STSA arises from the ${\bm z}
\leftrightarrow {\bm y}$ anti-symmetric part of the integrand.  In
arriving at \eq{dsigmaq} from \eq{eq-cross section} we employed the
lowest-order (order-$\as$) spin-dependent part of the light-cone wave
function squared \peq{eq-polarized wavefn}, which happens to be
symmetric under the ${\bm z} \leftrightarrow {\bm y}$ interchange:
hence, in our case, to obtain a contribution to the STSA the
interaction with the target had to be ${\bm z} \leftrightarrow {\bm
  y}$ anti-symmetric. However, it is possible that higher-order
corrections to the light-cone wave function squared would lead to a
${\bm z} \leftrightarrow {\bm y}$ anti-symmetric contribution. (By the
'wave function corrections' we understand all the initial and final
state corrections with rapidities between the projectile and the
particle we tag on.)  In such case, the interaction with the target
need not be ${\bm z} \leftrightarrow {\bm y}$ anti-symmetric, and can
be mediated by the standard $C$-even exchange. To test whether such
scenario is feasible within the CGC/saturation perturbative framework
one has to calculate the higher order corrections to the
polarization-dependent light-cone wave function squared
\peq{eq-polarized wavefn}.  The corrections would need a generate a
relative complex phase between the corrected and uncorrected wave
functions, in agreement with the initial proposal of
\cite{Brodsky:2002cx,Collins:2002kn}. This can be accomplished in LCPT
if the corrections lead to an intermediate state, for which the
imaginary part of the energy denominator leads to a non-vanishing
polarization-dependent contribution to the scattering amplitude. An
example of such corrections in our case could be a modification of the
amplitude in \fig{qtoqGampl} resulting from a gluon exchange between
the outgoing quark and gluon formed in the projectile splitting.
Calculation of such diagrams appears to be rather complicated and is
beyond the scope of this work.  However, it potentially may give a
contribution comparable to the STSA resulting from \eq{dsigmaq}: the
latter consists of the order-$\as$ light-cone wave function squared,
convoluted with the target interaction resumming powers of $\as^2 \,
A^{1/3}$ and $\as \, Y$, with one extra power of $\as$ due to the
odderon exchange \peq{O_ave}.  Our contribution \peq{dsigmaq} is,
therefore, order-$\as^2$, if one assumes that $\as^2 \, A^{1/3} \sim
1$ and $\as \, Y \sim 1$, which is parametrically comparable to the
$C$-odd order-$\as^2$ light-cone wave function squared, interacting
with the target through a $C$-even order-one exchange.  An explicit
calculation is needed to explore this possibility and is left for
future work.


\section*{Acknowledgments}

The authors are much indebted to Mike Lisa for strongly encouraging
them to think about spin physics at RHIC. We would like to also thank
Mickey Chiu, Alex Kovner, and Genya Levin for stimulating discussions
on the subject, and Daniel Boer, John Collins, Jianwei Qiu, and Raju
Venugopalan for comments on the manuscript. M.D.S. would like to thank
the Department of Energy's Institute for Nuclear Theory at the
University of Washington for the hospitality and constructive
atmosphere that facilitated this work.

This research is sponsored in part by the U.S. Department of Energy
under Grant No. DE-SC0004286.


\appendix

\renewcommand{\theequation}{A\arabic{equation}}
 \setcounter{equation}{0}
 \section{Averaging the odderon amplitude over target wave function}
\label{A}

\begin{figure}[ht]
 \includegraphics[width=4cm]{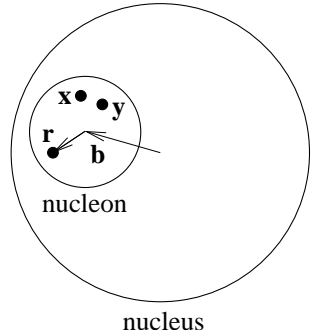}
 \caption{The geometry of the dipole--nucleus scattering as employed in \eq{Ocl1}.}
\label{odd_ave} 
\end{figure}

Let us construct the dipole odderon amplitude averaged over the target
field.  The triple gluon exchange happens between the dipole and a
nucleon in the target, which, for simplicity we model as a valence
quark in a bag.
(The overall factor in front of the averaged odderon
amplitude should indeed depend on the details of the averaging;
however, we believe the coordinate-space dependence would remain the
same for other models of the nuclear wave function.) The target
averaging then consists of averaging over the positions of the quark
in the nucleon and over the positions of nucleons in the nucleus,
along with summation over all nucleons. Assuming, again for
simplicity, that the quark has equal probability to be anywhere inside
the nucleon in the transverse plane (a cylindrical ``nucleon''
approximation), we write for the averaged odderon amplitude
\begin{eqnarray}
  \label{Ocl1}
  O_{\bm x \, \bm y} = c_0 \, \alpha_s^3 \, \int d^2 b \ T({\bm b}) \, 
\int \frac{d^2 r}{\pi \, a^2} \, \ln^3 
 \frac{|\bm x - \bm b - \bm r|}{|\bm y - \bm b - \bm r|}  \, \theta (a-r) \, 
\theta \left( a - \left| \frac{\bm x + \bm y}{2} - \bm b \right| \right)  \notag \\
\times \, \exp \left[ -\frac{1}{4} \, |\bm x - \bm y|^2 \ Q_s^2 \! 
\left( \frac{\bm x + \bm y}{2} \right) \, \ln \frac{1}{|\bm x - \bm y| \, \Lambda} \right].
\end{eqnarray}
Here $\bm b$ is the position of the center of a nucleon in the
transverse plane with respect to the center of the nucleus, $\bm r$ is
the position of the valence quark in the nucleon, and $a$ is the
radius of the nucleon, as illustrated in \fig{odd_ave}. The two
theta-functions in \eq{Ocl1} insure that the valence quark and the
center of the ${\bm x}, {\bm y}$-dipole are both located inside the
nucleon in the transverse plane.
In our simple model of the collision the dipole has to hit the nucleon
directly in order to be able to interact with the quarks inside of it.
Since the dipole ${\bm x}, {\bm y}$ is perturbatively small, we
enforce this condition by demanding that only the center of the dipole
is inside the nucleon's transverse extent. 

To integrate over $\bm r$ in \eq{Ocl1} we notice that
\begin{equation}
  \label{eq:zero}
  \int d^2 r \, \ln^3 \frac{|\bm x - \bm b - \bm r|}{|\bm y - \bm b - \bm r|} = 0
\end{equation}
if the integration carries over the whole transverse plane. Using this
result we write
\begin{equation}
  \label{inverse}
  \int d^2 r \, \ln^3 \left( \frac{|\bm x - \bm b - \bm r|}{|\bm y - \bm b - \bm r|} \right)
\, \theta (a-r) = - \int d^2 r \, \ln^3 
\left( \frac{|\bm x - \bm b - \bm r|}{|\bm y - \bm b - \bm r|} \right) \, \theta (r-a). 
\end{equation}
To approximate the integral on the right-hand-side of \eq{inverse} we
expand its integrand in powers of $|\bm x - \bm b|/r$ and $|\bm y -
\bm b|/r$ to the first non-trivial (after integration) order, thus
obtaining
\begin{equation}
  \label{ln3int}
  \int d^2 r \, \ln^3 \left( \frac{|\bm x - \bm b - \bm r|}{|\bm y - \bm b - \bm r|} \right)
\, \theta (a-r) \approx \frac{3 \, \pi}{8 \, a^2} \, |\bm x - \bm y|^2 \ (\bm x - \bm y) 
\cdot (\bm x + \bm y - 2 \, \bm b).
\end{equation}

Substituting \eq{ln3int} back into \eq{Ocl1} and defining a new
integration variable
\begin{equation}
  \label{delta_def}
  {\bm {\tilde b}} = {\bm b} - \frac{\bm x + \bm y}{2}
\end{equation}
yields
\begin{eqnarray}
  \label{Ocl2}
  O_{\bm x \, \bm y} \approx - c_0 \, \alpha_s^3 \, \frac{3}{4 \, a^4} \, |\bm x - \bm y|^2 \, 
 \exp \left[ -\frac{1}{4} \, |\bm x - \bm y|^2 \ Q_s^2 \!
\left( \frac{\bm x + \bm y}{2} \right) \, \ln \frac{1}{|\bm x - \bm y| \, \Lambda} \right] \notag \\ \times \, 
(\bm x - \bm y) \cdot \int d^2 {\tilde b} \ {\bm {\tilde b}} \
T \! \left(\frac{\bm x + \bm y}{2} + {\bm {\tilde b}} \right) \, 
   \theta \left( a - {\tilde b}_T \right) .
\end{eqnarray}

In principle this result is as far as one can simplify $O_{\bm x \,
  \bm y}$ without the explicit knowledge of the nuclear profile
function $T ({\bm b})$. To obtain a closed expression for the STSA in
the text we expand
\begin{equation}
  \label{expansion}
  T \! \left(\frac{\bm x + \bm y}{2} + {\bm {\tilde b}} \right) = 
T \! \left(\frac{\bm x + \bm y}{2} \right) + {\bm {\tilde b}} \cdot {\bm \nabla} 
T \! \left(\frac{\bm x + \bm y}{2} \right) + \ldots 
\end{equation}
with $\bm \nabla$ the transverse gradient operator.  Such expansion is
potentially dangerous near the edge of the nucleus profile, where the
derivatives may get large. For instance, for a solid-sphere model of
the nucleus the nuclear profile function is $T ({\bm b}) = \rho \, 2
\, \sqrt{R^2 - b^2}$ with $\rho$ the nucleon density and $R$ the
nuclear radius; the derivatives of such $T ({\bm b})$ near $b=R$ are
divergent. Using the realistic Woods-Saxon profile would make the
derivatives finite, but they would still be large. Thus we will
proceed by using the expansion \peq{expansion} as a way to simplify
the expression, keeping in mind that in the cases where this expansion
breaks down one has to return back to \eq{Ocl2}.

Substituting \eq{expansion} into \eq{Ocl2} and integrating over $\bm
{\tilde b}$ yields (for the first non-trivial term after integration)
\begin{equation}
  \label{Ocl3}
  O_{\bm x \, \bm y} \approx - c_0 \, \alpha_s^3 \, 
 \frac{3 \, \pi}{16} \, |\bm x - \bm y|^2 \, 
 \exp \left[ -\frac{1}{4} \, |\bm x - \bm y|^2 \ Q_s^2 \!
\left( \frac{\bm x + \bm y}{2} \right) \, 
\ln \frac{1}{|\bm x - \bm y| \, \Lambda} \right] \ (\bm x - \bm y) \cdot {\bm \nabla} 
T \! \left(\frac{\bm x + \bm y}{2} \right).
\end{equation}
This is our final expression for the target-averaged odderon
amplitude. Note an interesting feature of \eq{Ocl3}: the non-zero
contribution to the odderon amplitude in the transverse coordinate
space arises from the gradient of the nuclear profile function. The
odderon interaction with the target is thus only possible if the
target has a non-uniform profile in the transverse space. This is in
stark contrast to the $C$-even exchanges, which are non-zero even for
the Bjorken model of a nucleus of infinite transverse extent with
constant density in the transverse plane.



\providecommand{\href}[2]{#2}\begingroup\raggedright\endgroup


\end{document}